\documentclass{emulateapj}
\usepackage{epstopdf}
\usepackage{graphicx}
\newcommand{\beq}	{\begin{equation}}
\newcommand{\eeq}	{\end{equation}}
\newcommand{\beqa}{\begin{eqnarray}}
\newcommand{\eeqa}{\end{eqnarray}}

\def\simlt{\lower.5ex\hbox{$\; \buildrel < \over \sim \;$}}
\def\simgt{\lower.5ex\hbox{$\;. \buildrel > \over \sim \;$}}

\font\tenbi=cmmib10 
\newfam\bifam  \textfont\bifam=\tenbi
\font\tenbr=cmbx10
\newfam\brfam \def\br{\fam\brfam\tenbr} \textfont\brfam=\tenbr

\font\squinttenbi=cmbx10 at 9pt
\scriptfont\brfam=\squinttenbi

\def\vectimes{{\br{\times}}}
\def\vecnabla{
              \setbox1=\hbox{$\bigtriangledown$}
                           \raise.45ex\hbox{$\bigtriangledown$\hskip-.97\wd1
                           $\bigtriangledown$\hskip-.97\wd1
                           $\bigtriangledown$\hskip-.97\wd1}
                           \raise.47ex\hbox{$\bigtriangledown$}}

\def\curl{\vecnabla\vectimes}
\def\div{\vecnabla\cdot}
\def\rsun{\ifmmode {\rm R}_{\mathord\odot}\else $R_{\mathord\odot}$\fi}
\def\msun{\ifmmode {\rm M}_{\mathord\odot}\else $M_{\mathord\odot}$\fi}
\def\lsun{\ifmmode {\rm L}_{\mathord\odot}\else $L_{\mathord\odot}$\fi}

\newcommand{\kms}	{{\rm km}\, {\rm s}^{-1}}

\def\tmb{\ifmmode {T_{\rm mb}^{13}(x,y,v)}\else $T_{\rm mb}^{13}(x,y,v)$\fi}

\begin{document}

\title{Investigations of Protostellar Outflow Launching and Gas Entrainment: Hydrodynamic Simulations and Molecular Emission}

\author{Stella S.~R.~Offner\footnote{Hubble Fellow}}
\affil{Department of Astronomy, Yale University, New Haven, CT 06511}
\email{stella.offner@yale.edu }

\author{H\'ector G.~Arce}
\affil{Department of Astronomy, Yale University, New Haven, CT 06511}

\begin{abstract}
We investigate protostellar outflow evolution, gas entrainment, and star formation efficiency using radiation-hydrodynamic simulations of isolated, turbulent low-mass cores. We adopt an X-wind launching model, in which the outflow rate is coupled to the instantaneous protostellar accretion rate and evolution. We vary the outflow collimation angle from $\theta=0.01-0.1$ and find that even well collimated outflows effectively sweep up and entrain significant core mass. The Stage 0 lifetime ranges from 0.14-0.19 Myr, which is similar to the observed Class 0 lifetime. The star formation efficiency of the cores spans 0.41-0.51.  In all cases, the outflows drive strong turbulence in the surrounding material. Although the initial core turbulence is purely solenoidal by construction, the simulations converge to approximate equipartition between solenoidal and compressive motions due to a combination of outflow driving and collapse. When compared to simulation of a cluster of protostars, which is not gravitationally centrally condensed, we find that the outflows drive motions that are mainly solenoidal. The final turbulent velocity dispersion is about twice the initial value of the cores, indicating that an individual outflow is easily able to replenish turbulent motions on sub-parsec scales.  We post-process the simulations to produce synthetic molecular line emission maps of $^{12}$CO, $^{13}$CO, and C$^{18}$O and evaluate how well these tracers reproduce the underlying mass and velocity structure. %compare these to the observed morphologies and properties of outflows. % recently observed with ALMA.
\end{abstract}
%Check what the rms velocity is for the different tracers

\section{Introduction}

During formation, stars launch high-velocity, collimated mass outflows that impact the local gas and global cloud environment. Since the complex physics of outflow launching occurs on sub-AU scales and happens during the protostellar phase while the forming star is enshrouded by dense dust and gas, the process is observationally difficult to investigate.  High-resolution observations by ALMA, which can probe AU scales may shed light on this process \citep{arce13}, but current observational data fail to probe the appropriate scales.  Consequently, the details of the ``central engine'' that powers outflows is not well understood. 

Mass outflows from forming stars bear on a number of fundamental open questions in star formation. By entraining and unbinding infalling material, outflows reduce the efficiency at which gas accretes onto protostars \citep{matzner00,arce02,arce05}. This has implications for the stellar characteristic mass and stellar initial mass function \citep{krumholz12}. Outflow activity may also partially explain the difference between the distribution of prestellar core masses and stellar masses, which  differ by a factor of $\sim1/3$ \citep{enoch07,alves07,rathborne09}. All molecular clouds are observed to be turbulent, but the origin of this turbulent energy is unknown. Protostellar outflows are one mechanism that may replenish turbulent motions \citep{bally99,swift08,nakamura11,nakamura07,carroll09, cunningham09, carroll10, wang10,li10,hansen12}.

Several analytic theories have been proposed to explain outflow characteristics \citep{shu94,pelletier92,li&shu96,matzner99}.  These theories are based on the strong coupling between magnetic fields and accreting gas. As a result of the subsequent magnetic enhancement  and field winding during the collapse process, mass is ejected at high-velocities along the field lines. 

%Duffin & pudritz have 6.5 AU resolution
%Lovelace et al. 2010, 2.5 D, doesn't say resolution, no-non ideal effects
Numerical simulations provide an alternative avenue to explore outflow launching and characteristics. However, the complexity of the physics and the wide range of scales involved ($\sim$0.01 AU-1 pc) prohibit a first-principles approach in all aspects of the problem. Consequently numerical simulations focus on one of three regimes. Studies investigating the outflow engine model $\lesssim$ 100 AU scales with sub-AU resolution 
% include non-ideal magnetic effects in order to capture magnetic dissipation 
\citep{lovelace10,lii12,price12,tomida13}. Calculations focusing on the formation of an individual star and role of outflow activity model larger volumes at lower resolution, e.g., $dx >$ 1AU \citep{duffin09,seifried11,machida13}. In some cases, models may adopt a simplified model for the outflow launching \citep{lee00,lee01,rosen04,banerjee07,Offner11}. Some of these calculations are able to follow the launching of outflows self-consistently, but generally resolution is not sufficiently high to create a well collimated jet on sub-AU scales (e.g., \citealt{duffin09,seifried11,machida13}) and calculations may be too computationally expensive to follow for long evolutionary times (e.g., \citealt{price12,tomida13}).  %Some of these studies are able to explai
Finally, studies investigating the interaction of outflows with the parent cloud and subsequent impact on cluster properties focus on $\sim$ pc scales and adopt a sub-grid model for outflow launching \citep{nakamura07, carroll09, carroll10,offner12,hansen12,wang10,li10,krumholz12}. Such models are generally motivated by previous analytic work and coupled to hydrodynamic quantities such as the stellar mass and accretion rate. % to be more self-consistent.

In this paper we focus on the intermediate scales ($\sim$20 AU -0.2pc) on which outflows interact with their local environment. We perform numerical simulations of the collapse of turbulent, dense low-mass cores. We adopt a sub-grid model for the outflow launching, which allows us to vary the outflow opening angle and investigate the impact on protostellar accretion and evolution. In section \ref{methods} we describe the simulation methodology. In section \ref{results} we describe the simulation outcomes, including the distribution of gas and protostellar properties. We then post-process the simulations to obtain molecular line emission %and compare with observations 
in section \ref{obs}. We summarize our results in section \ref{summary}.

\section{Numerical Methods}\label{methods}

%For comparison Cunningham 2011 runs for 0.8 tff (tff=50.7, 30.2 kyr), maximum res=25 AU

We perform the simulations using the ORION Adaptive Mesh Refinement (AMR) code \citep{klein99, truelove98}. We include self-gravity and compute the gas temperature by solving the equations of radiative transfer in the Flux-Limited Diffusion (FLD) approximation \citep{krumholzkmb07}. This treatment assumes the gas and dust are perfectly thermally coupled, and we adopt a standard solar metallicity dust composition \citep{semenov03}. In very hot gas, the dust sublimates and temperatures can no longer be computed with the FLD solver. Instead, we include the treatment for atomic line cooling described by \citet{cunningham11}, which implicitly solves for gas temperatures exceeding $10^4$ K.
%which assumes a standard iron abundance and treats the grains as composite aggregates. The strong bow shocks produced by outflowing gas running into ambient material can generate temperatures well in excess of the dust destruction temperature. Since the outflows simulated here are young and very low mass, only a small number of cells ever reach temperatures above 1000 K. Nonetheless, we include the treatment for atomic line cooling described by Cunningham et al. (2011), which implicitly solves for cell temperatures exceeding 104 K.

We initialize the calculations with a 10 K sphere of radius $L=0.064$ pc and  uniform density $\rho_{\rm core} = 2\times10^{-19}$ g cm$^{-3}$, which corresponds to a total mass of 4 $\msun$. This cold core is pressure confined by a hot, low-density medium. The velocity field of the cold, dense sphere is initialized with a random field of perturbations having Fourier modes $1 \le k \le 2 $. The Mach number of the gas, $\mathcal{M}= 2.5$,  is chosen so that the core satisfies the linewidth-size relation \citep{MandO07}. The calculations use a basegrid of 64$^3$, where the sphere is resolved to level 2, and we insert five or seven AMR levels such that the minimum cell size is $\Delta x_{\rm min}=$26 or 6.5 AU.  Additional refinement occurs if either of three criteria is met. To avoid artificial gravitational fragmentation, the grid is refined if the gas density exceeds the local Jeans density, $\rho > \rho_J \equiv N_J^2 \pi c_{s}^2/( G  \Delta x^2)$, where $c_s$ is the local sound speed and $N_J=0.125$ is the Jeans number \citep{truelove97}. Any gas with density $ \rho> 0.5 \rho_{\rm core}$ is always refined to level 2 ($\Delta x =210\,$AU) whether or not it exceeds the local Jeans density. These two criteria ensure that the outflow cavity walls and entrained material are always followed at higher resolution. Finally, higher resolution is added if there is a strong gradient in the radiation energy density: $| \nabla E_{rad}|/E_{rad} > 2$. This criterion facilitates the convergence of the radiation solver and forces cells in the radiatively heated region near the protostar to be refined even if the gas densities are low. Protostars are always contained within $16^3$ cells on the finest level, which is where the accretion and wind launching takes place. We adopt outflow boundary conditions so that outflowing material can freely escape the domain.  The simulation properties are summarized in Table \ref{simprop}.

As the calculation proceeds the initial turbulence decays, allowing the gas to gravitationally collapse. When the Jeans density is exceeded on the maximum level a star particle is inserted \citep{krumholz04}. These particles represent individual protostars and follow a sub-grid stellar evolution model that includes accretion luminosity down to the stellar surface, Kelvin-Helmholz contraction, and nuclear burning \citep{Offner09}.  The protostars are also endowed with a model for the launching of protostellar outflows based upon \citep{matzner99}. This model has previously  been used in other outflow studies  \citep{cunningham11,Offner11,krumholz12,hansen12,offner12}.  The model is characterized by three dimensionless parameters that specify the outflow ejection efficiency, outflow velocity, and momentum distribution. The mass ejection rate, $f_w$, gives the fraction of accreting gas that is launched in the wind. This fraction is observationally uncertain (e.g., \citealt{plunkett13}), but the disk wind (\citet{pelletier92} and X-wind \citep{shu88} models predict $f_w=$ 0.1-0.33. Here, we adopt $f_w$ = 0.2-0.3. Consequently, $1.0/(1 + f_w)$ of the infalling gas accretes onto the star, while $f_w/(1 + f_w)$ is launched in an outflow.

For the outflow launching velocity, we adopt the Keplerian velocity near the stellar surface, $v_{w} = f_v\sqrt{ GM_*/r_*}$, where $M_*$ is the protostellar mass, $r_*$ is the protostellar radius, and $f_v$ is a model parameter relating to the wind acceleration.  The X-wind and disk wind models both predict $f_w f_v \sim 0.3$, while surveys of outflows suggest a range of values, $0.025 \lesssim f_w f_v \lesssim 0.38$, with no apparent dependence on spectral class \citep{cunningham11}. Here we adopt $f_v = 0.33$ such that $f_w f_v \simeq 0.07-0.1$. In this model, the launch velocity increases fairly strongly with mass, and so the outflow speed increases over the course of the simulation. We set the wind temperature to $10^4$ K, the temperature of an ionized wind.
%, greatly constraining the numerical time step of the calculation. Here, we follow \citet{cunningham11} by setting the launching velocity to a fraction of Keplerian velocity, i.e., $\sqrt{3}$ times lower, which alternatively corresponds to a launching radius that is three times the protostellar radius. Although the details of outflow launching are not well understood, it is likely that mass is ejected from a range of radii near the protostar. Various observed outflows have a range of outflow speeds and collimations, which are variously consistent with the X-wind and disk wind models \citep{arce07}.

The outflow direction is set by the direction of the angular momentum vector of the protostar. This is in turn determined by the time-integrated angular momentum of the accreting gas, a quantity that depends upon the turbulent properties of the core and evolves over the calculation. Thus, the wind direction is not fixed but is self-consistently dictated by the hydrodynamic evolution of the accretion flow. However, we do not find large variations in the outflow direction or precession, which has been observed in some outflows \citep{ybarra06,wu09}. The outflow axis is generally parallel to the angular momentum vector of the accretion disk.

We define the effective opening angle of the wind, $\theta_0$, following \citet{matzner99}. They write the angular momentum distribution of the outflow in terms of the polar angle measured from the protostar's rotational axis $\theta_0$: 
\begin{equation} 
\xi (\theta, \theta _{\rm 0})= \left[{\rm ln} \left(\frac{2}{\theta _{\rm 0}} \right)\big({\rm sin}^2\theta + \theta _{\rm 0}^2\big) \right]^{-1}.
\end{equation}
Based on observed low-mass protostars, \citet{matzner99} infer that $\theta_0 \lesssim 0.05$ and suggest a fiducial value of $\theta_0$ = 0.01, which produces strong collimation. Here, we adopt $\theta_0=$ 0.01, 0.03, and 0.1. Small values of $\theta_0$ result in most of the outflow momentum being deposited in the grid within a few cells along the rotational axes.
	
We inject the wind into the refined cells with radial distances $4\Delta x < r \le 8 \Delta x$ from the protostar. By construction, the injection region is exterior to the accretion region, so that the wind does not  impede accretion along the disk midplane.
%In practice, $\bar{\xi }$ is also set to 0 when θ becomes close to π/2. Our outflow algorithm is fully mass conserving. Gas that was previously automatically accreted onto the stars in the original simulation is instead divided between accreted gas and outflow gas that is deposited back onto the numerical grid.
%

We evolve each of the calculations for 0.5 Myr, which corresponds to approximately four core free-fall times. Observationally, the embedded phase lasts $\sim 0.4$ Myr \citep{evans09}. Thus, we follow the core evolution until most of the initial core mass is either accreted or expelled by the outflow.

At the final time, the calculations each have a single protostar. Additional protostars may form during the evolution, but protostars that approach within four cells of the primary and also have a mass $m_* < 0.1 \msun$ are merged. We find that one or more additional protostars do form in our studies, generally in the early stage of collapse before the formation of an organized Keplerian disk around the primary, but their masses remain below $ 0.1\msun$ and they don't survive as independent protostars. 

To check for convergence, we perform an additional calculation, th0.1fw0.3h, with 7 AMR levels and $\Delta x_{\rm min}=6.5$ AU, which otherwise has identical model parameters to th0.1fw0.3 (see Table \ref{simprop}). Model th0.1fw0.3h produces one fewer fragment at early times and the fragment masses are smaller than those in model th0.1. This leads us to conclude that higher resolution models would converge to a single or, at most a close, binary system.

\begin{figure}
%\epsscale{2.25}
%\plottwo{plrc00640_Slice_z_Density.eps}{pla700800_Slice_z_Density.eps}
\epsscale{1.2}
\plotone{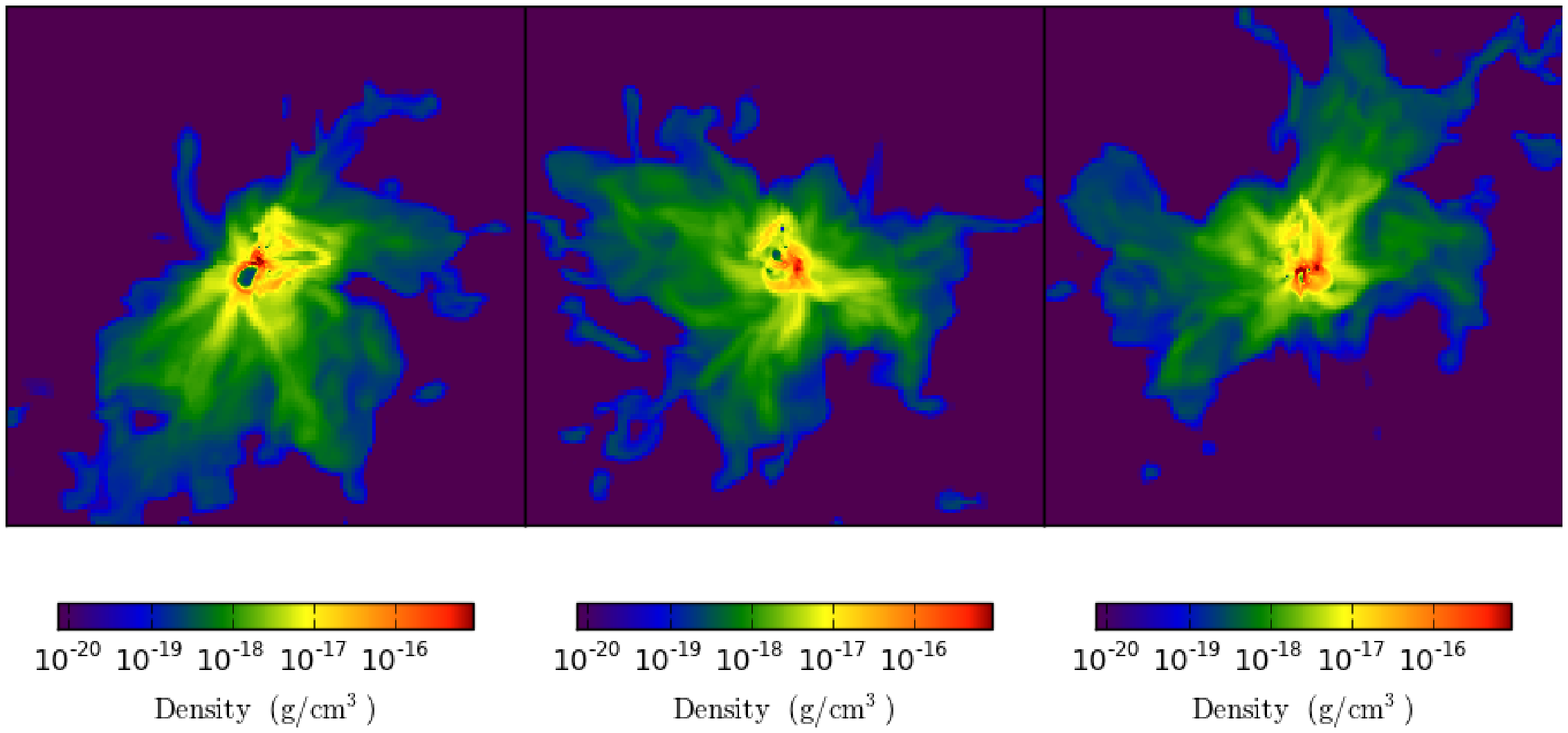}
\vspace{-0.55in}
\plotone{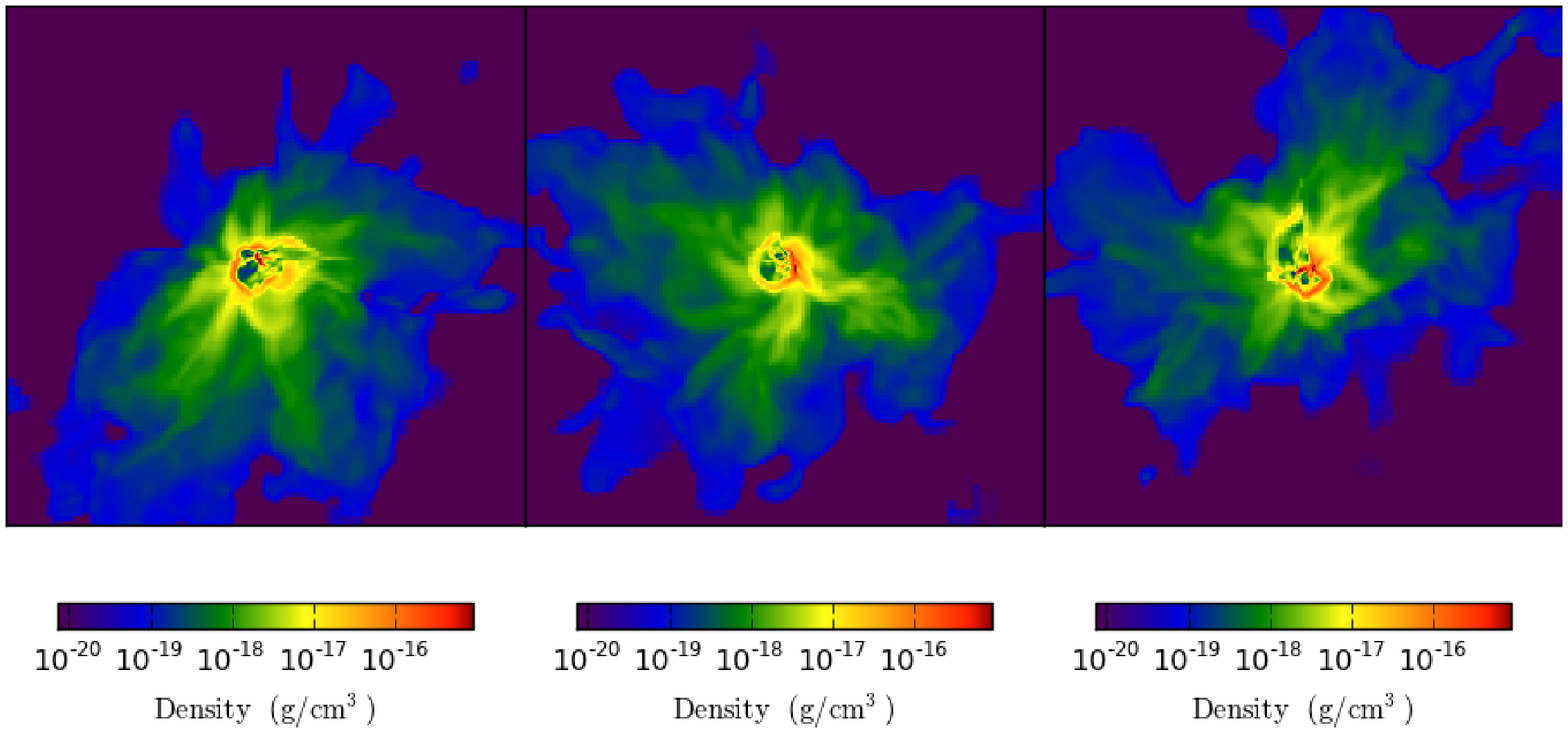}
\caption{A slice of log density through the center in the $x$, $y$ and $z$ planes for models th0.1fw0.3 (top) and th0.1fw0.3h (bottom) at 0.19 Myr.
\label{krho0_slice_conv} }
\end{figure}

\begin{figure}
\epsscale{1.2}
\plotone{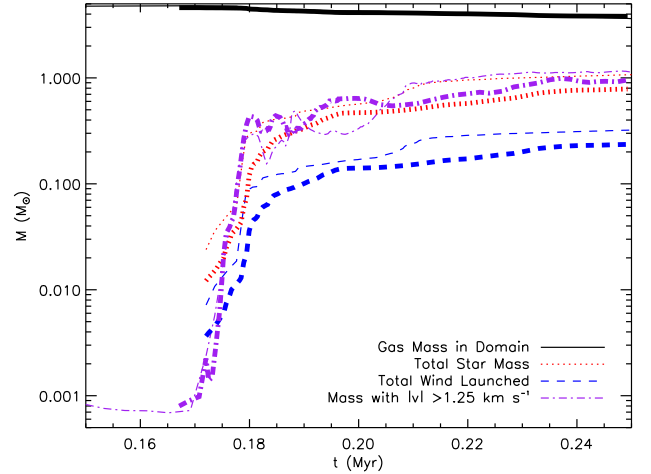}
\caption{Domain mass, star mass, launched mass, and mass with $|v|>0.1$ km s$^{-1}$ as a function of time for runs th0.1fw0.3 (thin) and th0.1fw0.3h (thick).
\label{acc_cont} }
\end{figure}

Figure \ref{krho0_slice_conv} shows the density distribution for models th0.1fw0.3 and th0.1fw0.3h, which are very similar. Figure \ref{acc_cont} shows the protostellar masses, wind launched, domain gas mass and mass of high-velocity gas as a function of time. The trends are very similar, although the protostar in th0.1fw0.3 is formed earlier with a slightly higher mass.  The outflow launching and protostellar luminosity depend upon the protostellar radius, which in turn depends upon the interior stellar state.  A slight difference in mass in the two cases leads to one protostar progressing to deuterium burning before the other, which impacts the protostellar radius and introduces small differences between gas velocities. The domain mass evolution in the two runs is sufficiently similar that the black thin line for run th0.1fw0.3 is hidden beneath the th0.1fw0.3h line. 

In principle the amount of entrainment may be enhanced or suppressed depending on the simulation resolution. Run th0.1fw0.3h contains more cells at higher resolution and has a larger resolution gradient between the protostar and the ambient medium. However, Figure \ref{acc_cont} shows that the fraction of high velocity material, $v>1.25\kms$, is similar in the two runs, which suggests that gas entrainment is not strongly sensitive to resolution to the extent we vary it here. 

For completeness we also consider run cl.th0.01, which follows the formation of a  cluster of protostars and employs the same outflow launching model. The details of the simulation are described in \citet{offner12}. We use this simulation to contrast the turbulence generated from an ensemble of outflows with that of an isolated protostar.

%ytorionlib/calc_gas_mass.py
%Documents/lib/idlamarlib/outflow_prop
%Documents/lib/idlamarlib/plot_all_outflow_prop

\begin{deluxetable}{lccccccc}
\tablecolumns{5}
\tablecaption{Model Properties \label{simprop}}
\tablehead{ \colhead{Model\tablenotemark{a}} &  
  \colhead{$M$($\msun$)} &
  \colhead{$\Delta x_{\rm max}$(AU)} &
  \colhead{$\theta_0$(rad)} &
 \colhead{$f_{w}$} &
 \colhead{$t_f$(Myr)} } \\
\startdata
th0.1fw0.3     & 4.0 & 26  & 0.1 & 0.3 & 0.5  \\ %74
th0.1fw0.3h    & 4.0 & 6.5 & 0.1  & 0.3 & 0.3 \\ %74
th0.1       & 4.0 & 26  & 0.1 & 0.2 & 0.5  \\ %74
%krho0\_m0.01 & 4.0 & 26  & $\frac{1}{3}$ & 0 & 0.01 & 0.1 \\ %74
%krho1.5\_0.01     & 4.0 & 26  & $\frac{1}{3}$ & 1.5 & 0.01 & 0.1\\ %13
%th0.04      & 4.0 & 26  & 0.04 & 0.2 & 0.5 \\ 
th0.03      & 4.0 & 26  & 0.03 & 0.2 & 0.5 \\
th0.01      & 4.0 & 26  & 0.01 & 0.2 & 0.5 \\ 
cl.th0.01\tablenotemark{b}         & 180 & 128 & 0.01 & 0.3 & 0.3 \\ % Check properties
\enddata
\tablenotetext{a}{%Simulation ID, core mass, minimum cell size, launching velocity in units of the Keplerian velocity, initial density profile where $\rho(r) \propto r^{-\kappa_{\rho}}$, maximum sink particle merge mass and opening angle (as defined by Matzner \& McKee 99). 
The model name, gas mass, simulation resolution on the maximum level, effective opening angle, fraction of accreted mass ejected in a wind, and the final time of the simulation. The molecular cores have initial radii of 0.064pc, temperature 10 K, and turbulence given by the linewidth-size relation $v_{\rm rms}= 0.5$ km s$^{-1}$. }
\tablenotetext{b}{Turbulent clump simulation of a forming star cluster, which has $L=0.65$ pc, initial temperature 10 K, and  $v_{\rm rms}= 1$ km s$^{-1}$. See \citet{offner12} for additional details.}
\end{deluxetable}
%\bar rho =2.4 x

\begin{deluxetable*}{lcccccccc}
\tablecolumns{6}
\tablecaption{Model Outcomes \label{simsum}}
\tablehead{ \colhead{Model\tablenotemark{a}} &  
  \colhead{$M_{*,f}$($\msun$)} &
  \colhead{$M_{D,f}$($\msun$)} & 
 \colhead{$M_{100K,f}$($\msun$)} &
  \colhead{$\epsilon_*$} &
 \colhead{ $v_{\rm rms}$(km s$^{-1}$}) &
 \colhead {$t_{\rm 0}$} &
 \colhead{SFR$_{\rm ff}$}}\\
\startdata
th0.1fw0.3    & 1.45 & 1.3  & 0.9  & 0.47 (0.36) & 1.0 & 0.19 & 0.15 \\
th0.1        & 1.73 & 1.04 & 0.63 & 0.51 (0.43) & 1.3 & 0.15 & 0.18 \\
%th0.04     & 1.08 & 0.55 & 0.38 & 0.30 (0.27) & 5.8 \\
th0.03      & 1.54 & 0.48  & 0.27 & 0.41 (0.39)  &  1.2 & 0.17 & 0.16   \\ %0.34
th0.01     & 1.60 & 0.77 & 0.4 & 0.44 (0.40) & 2.2 & 0.14 & 0.16 \\
cl.th0.01     & 9.0 & 171 & 170 & 0.05 (0.05) & 1.5 & * & 0.05 \\
%th0.1h     &  0.80    & 3.77 &  3.65  &    (0.20) &     %time = 0.3 Myr
\enddata
\tablenotetext{a}{Model name, final stellar mass, final remaining gas on the domain, remaining gas that has $T<100$ K, the star formation efficiency $\frac{M_{*,f}}{4\msun - M_{{\rm 100 K}, f}}$ $\left( \frac{M_{*,f}}{4\msun}\right)$, the final 3D velocity dispersion, the Stage 0 lifetime, and the star formation rate per free fall time.}
\end{deluxetable*}

%Plot efficiency as a function of time: M*/(Mi-Mcold, core)
% Is Mach number of cold & all (hot + cold) very different?
% Check vmax for these
 
\section{Results}\label{results}

\subsection{Outflow Morphology}

\begin{figure*}
%\epsscale{1.1}
\plottwo{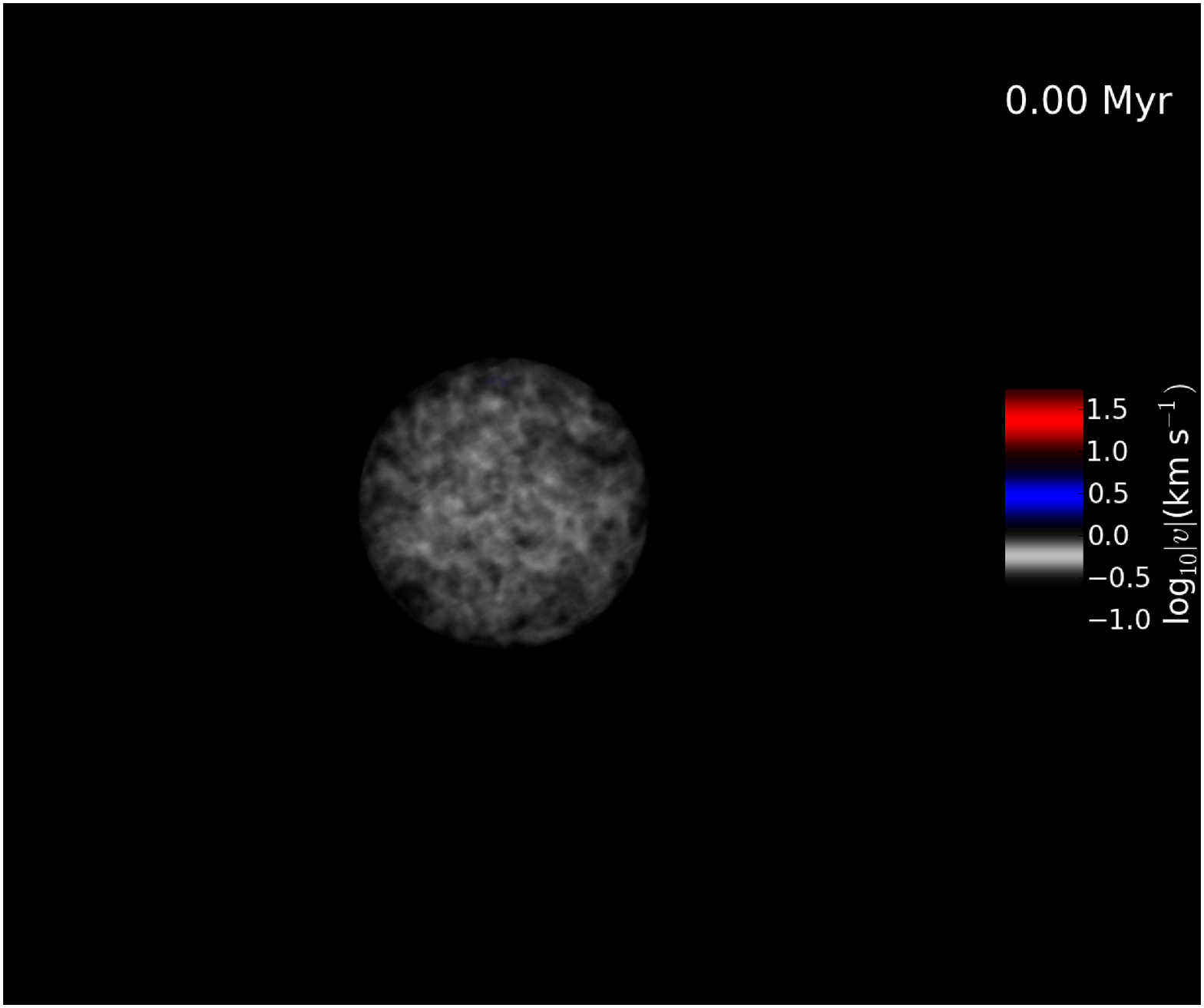}{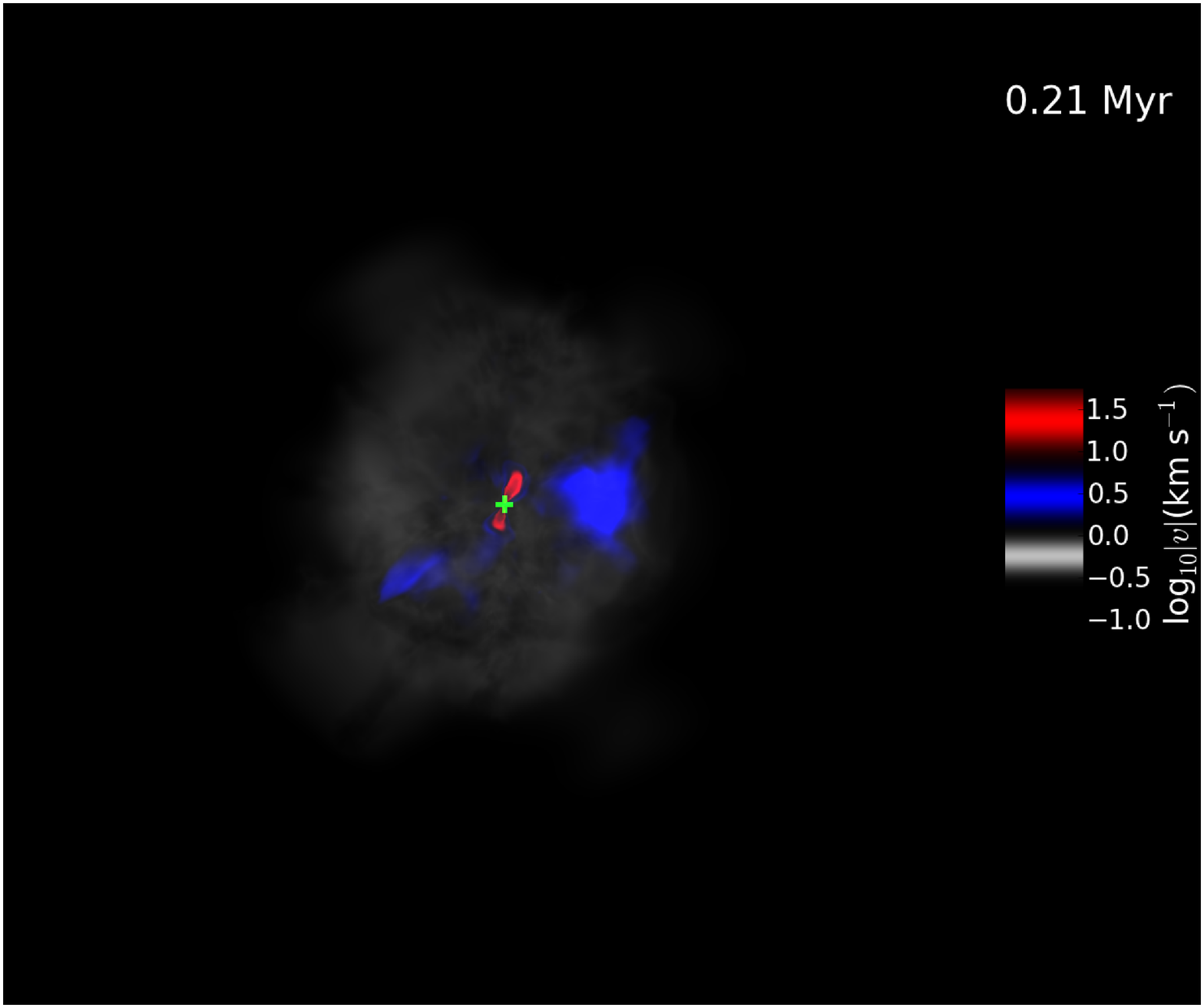}
%\epsscale{2.43}
\plottwo{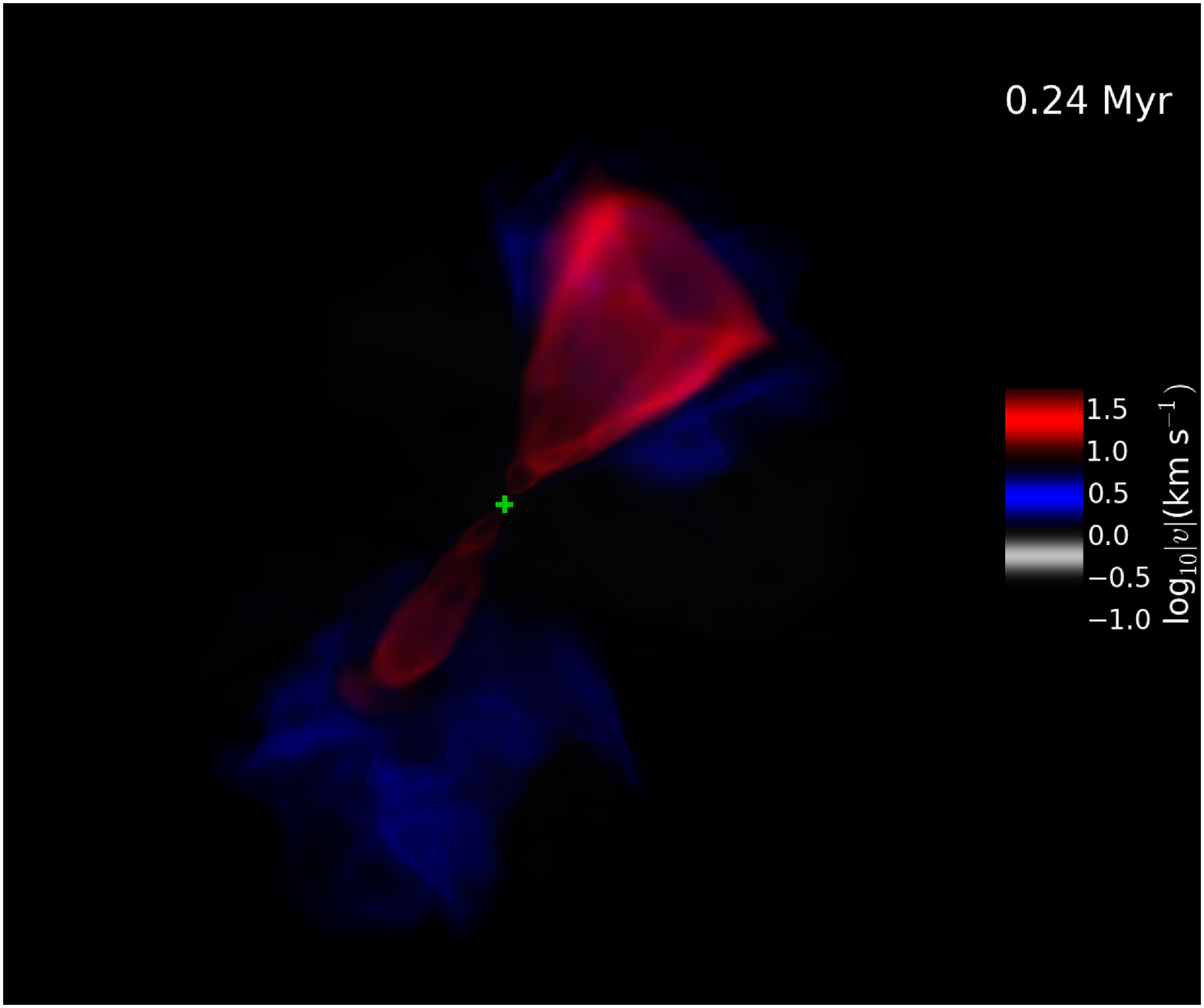}{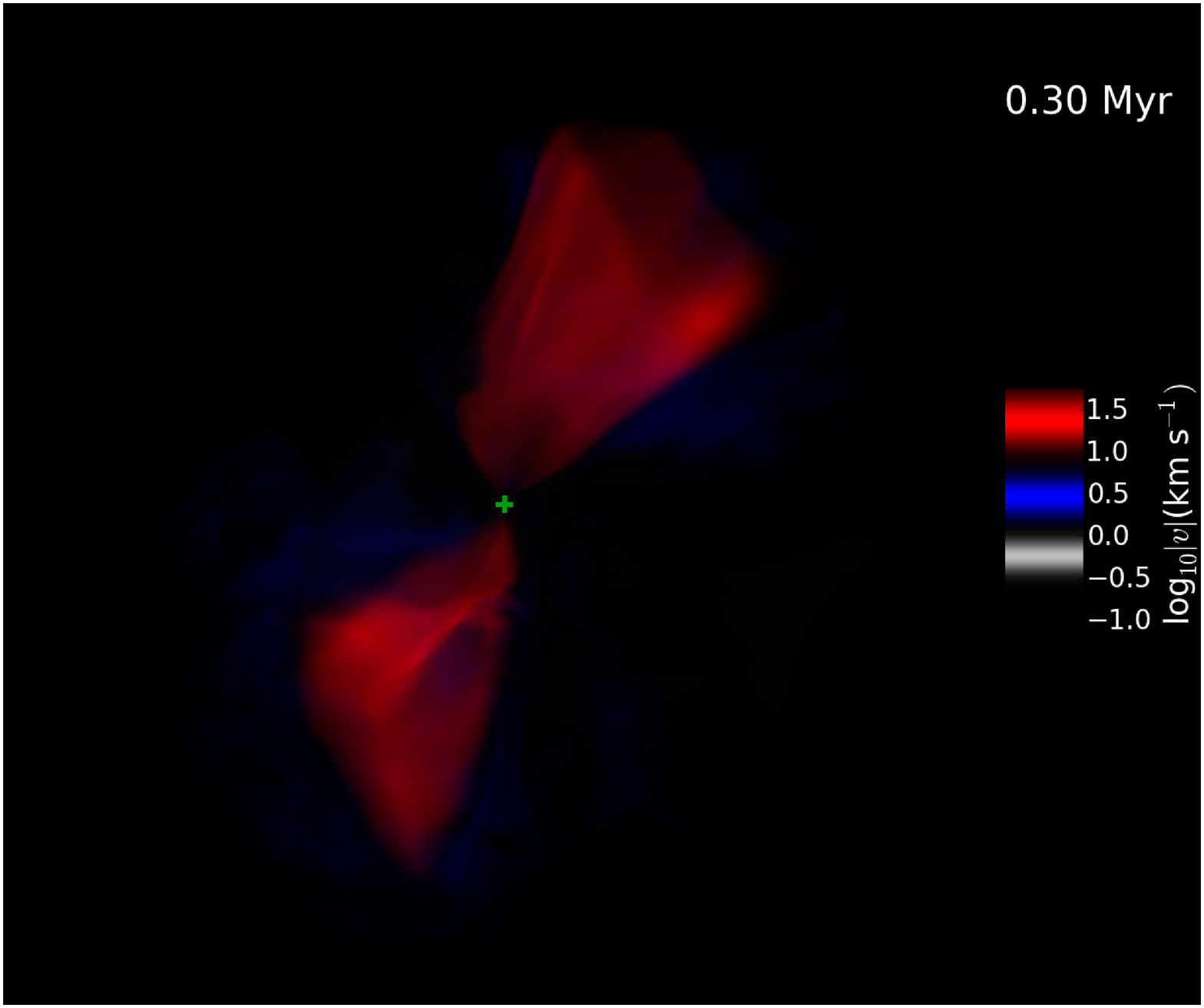}
%\epsscale{5.4}
\plottwo{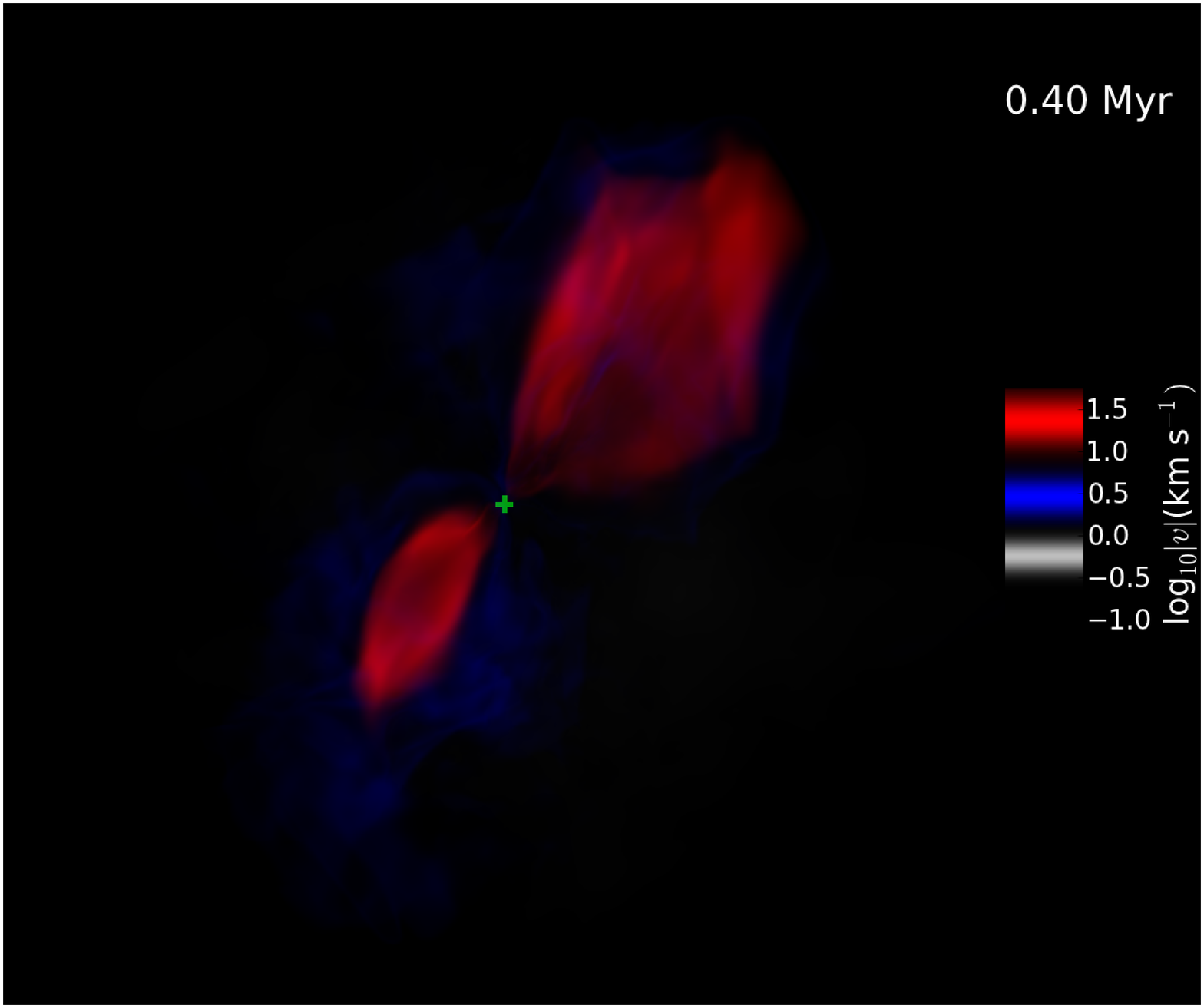}{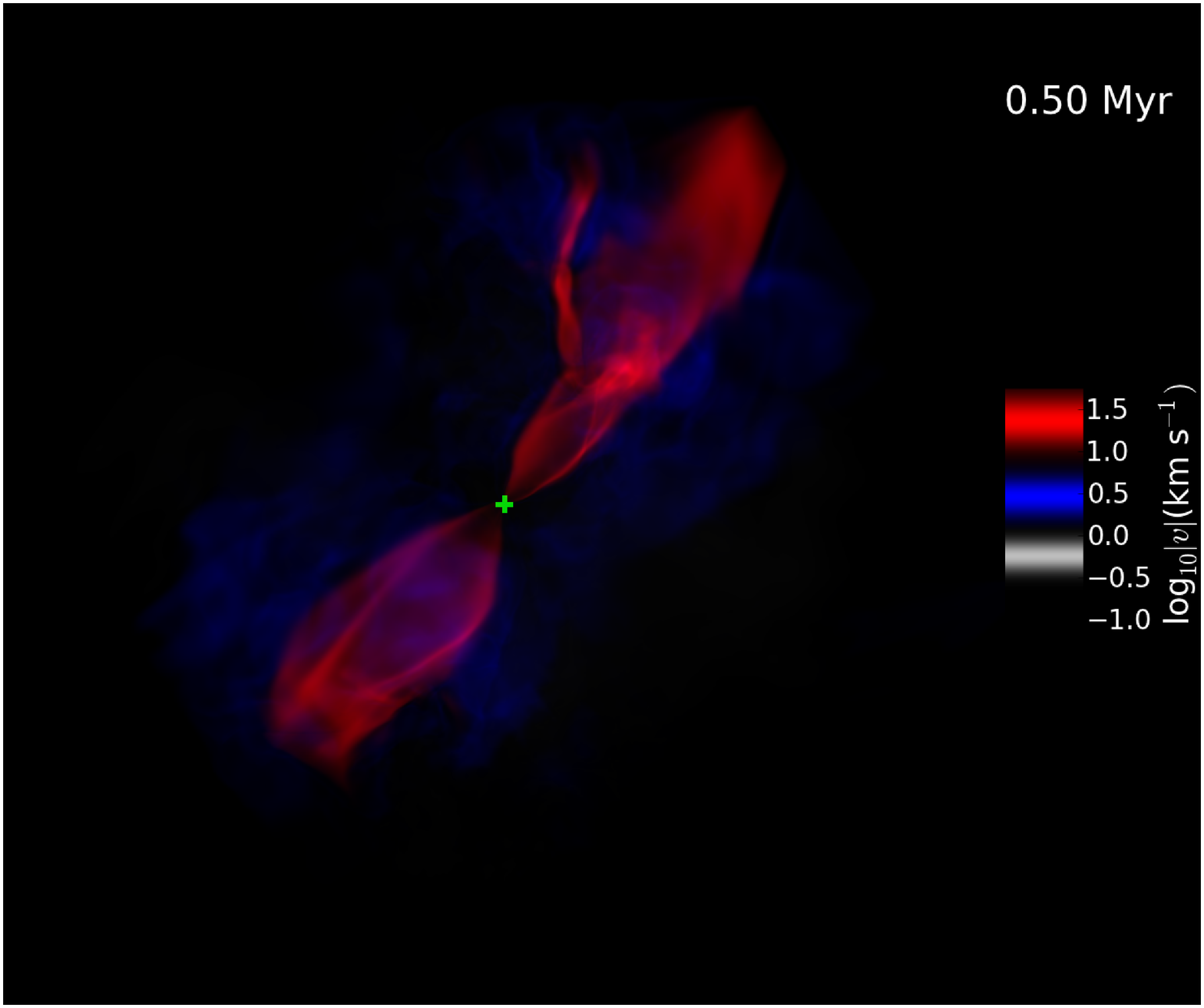}
\caption{Volume rendering of gas velocity for run th0.1fw0.3 at six different times.The box width is 0.5 pc. The location of the protostar is marked with a green cross. Times are indicated in the upper right. A movie is available showing the full time sequence.
\label{VR_1} }
\end{figure*}

\begin{figure*}
%\epsscale{1.1}
\plottwo{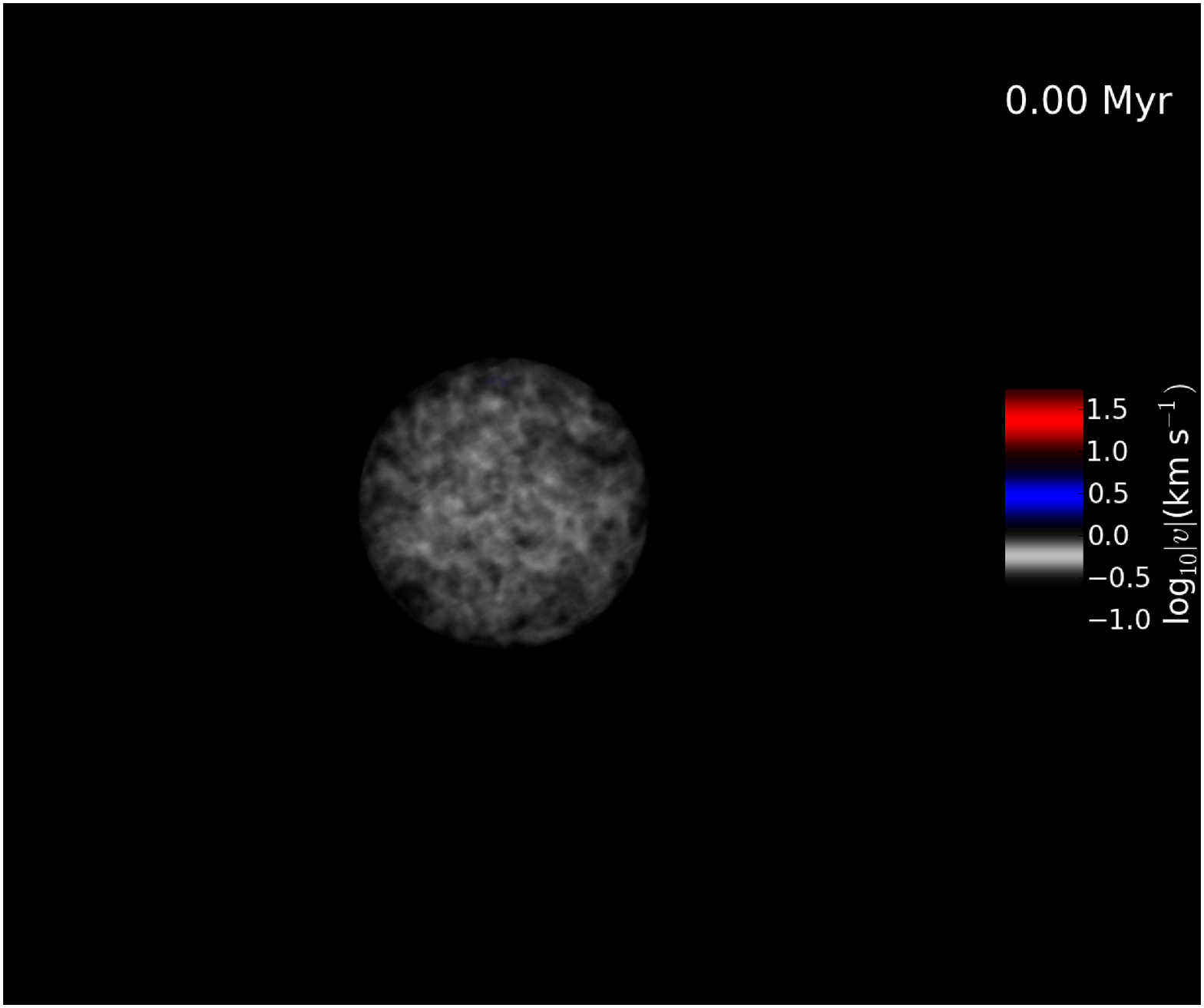}{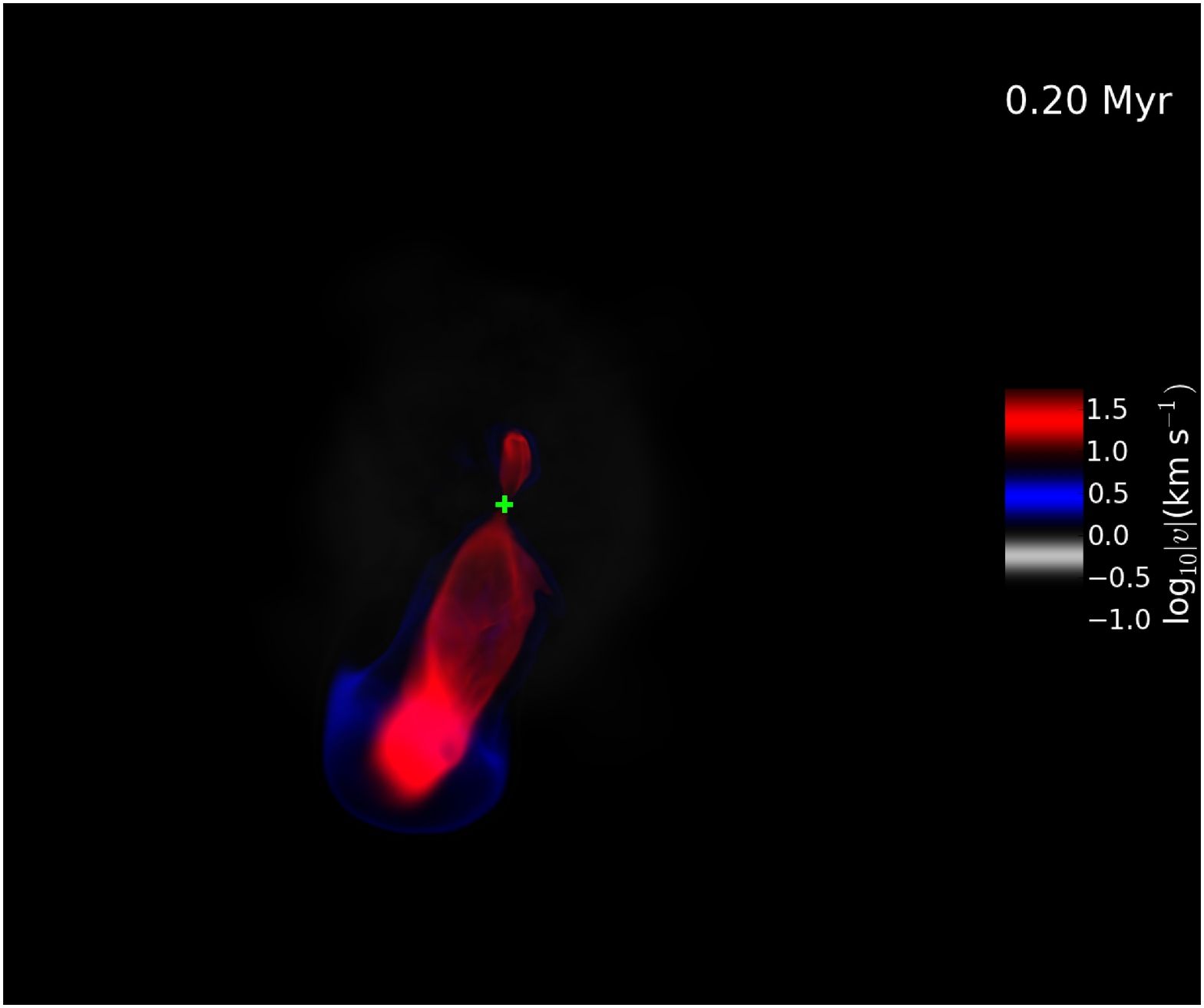}
%\epsscale{2.43}
\plottwo{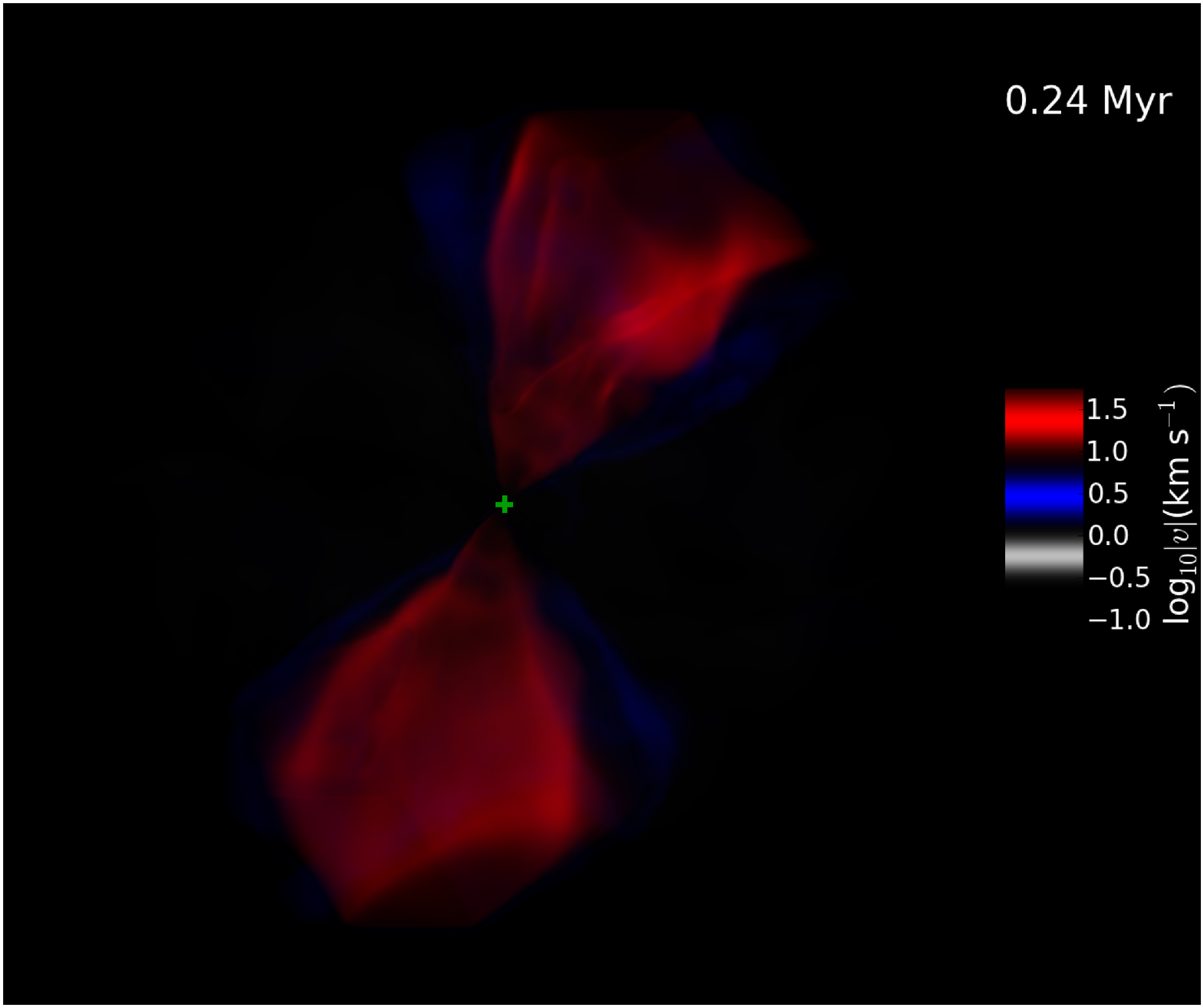}{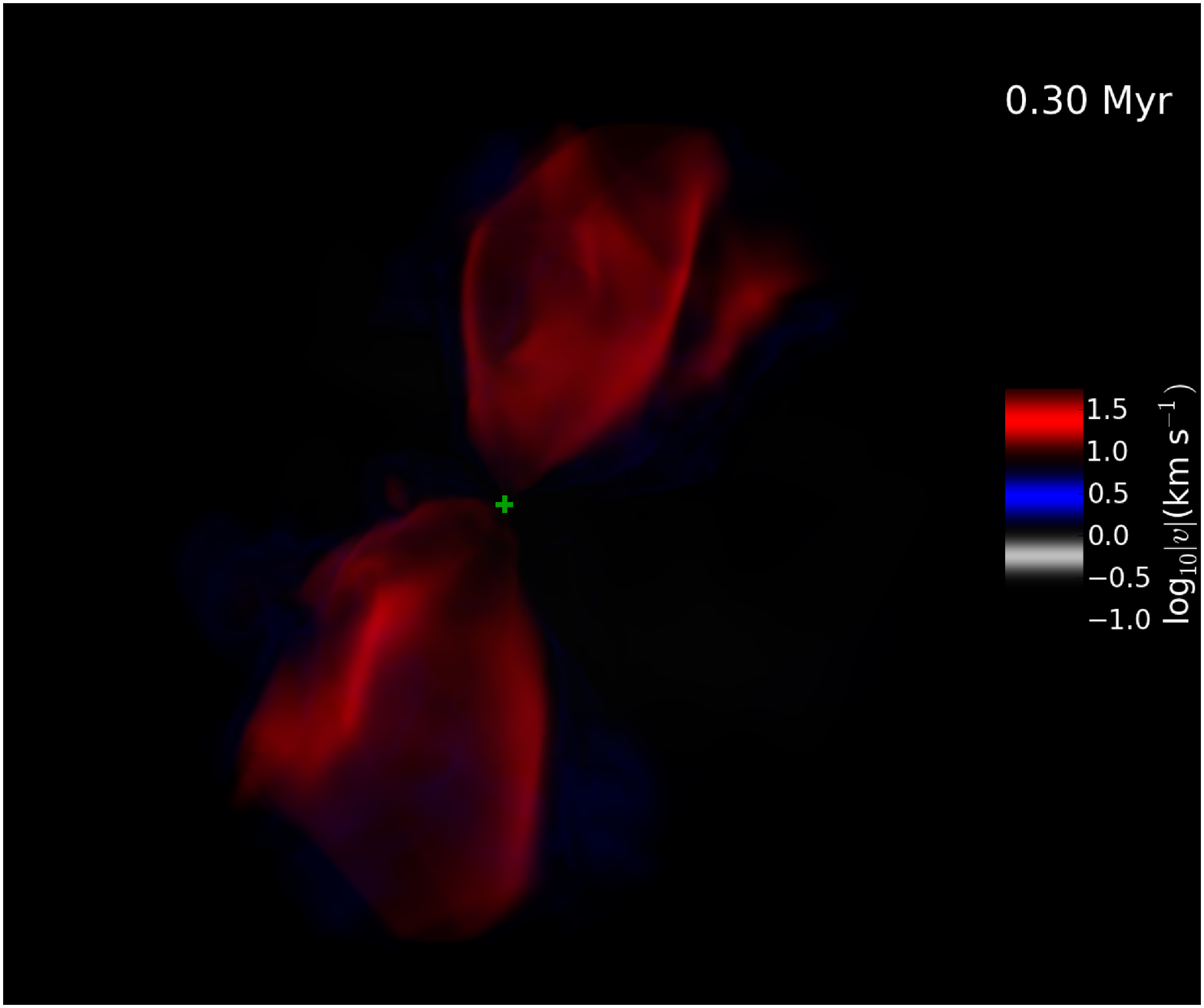}
%\epsscale{5.4}
\plottwo{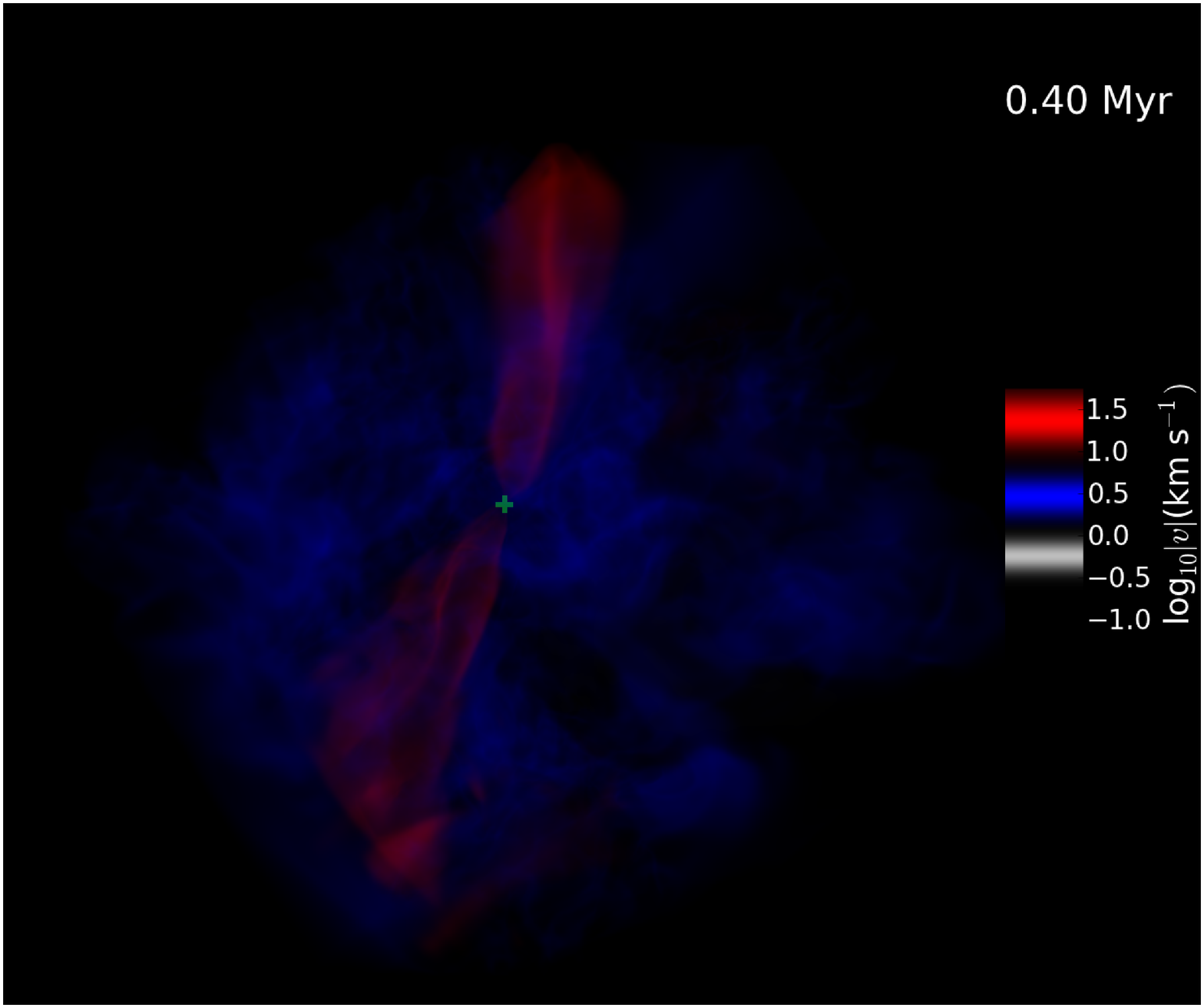}{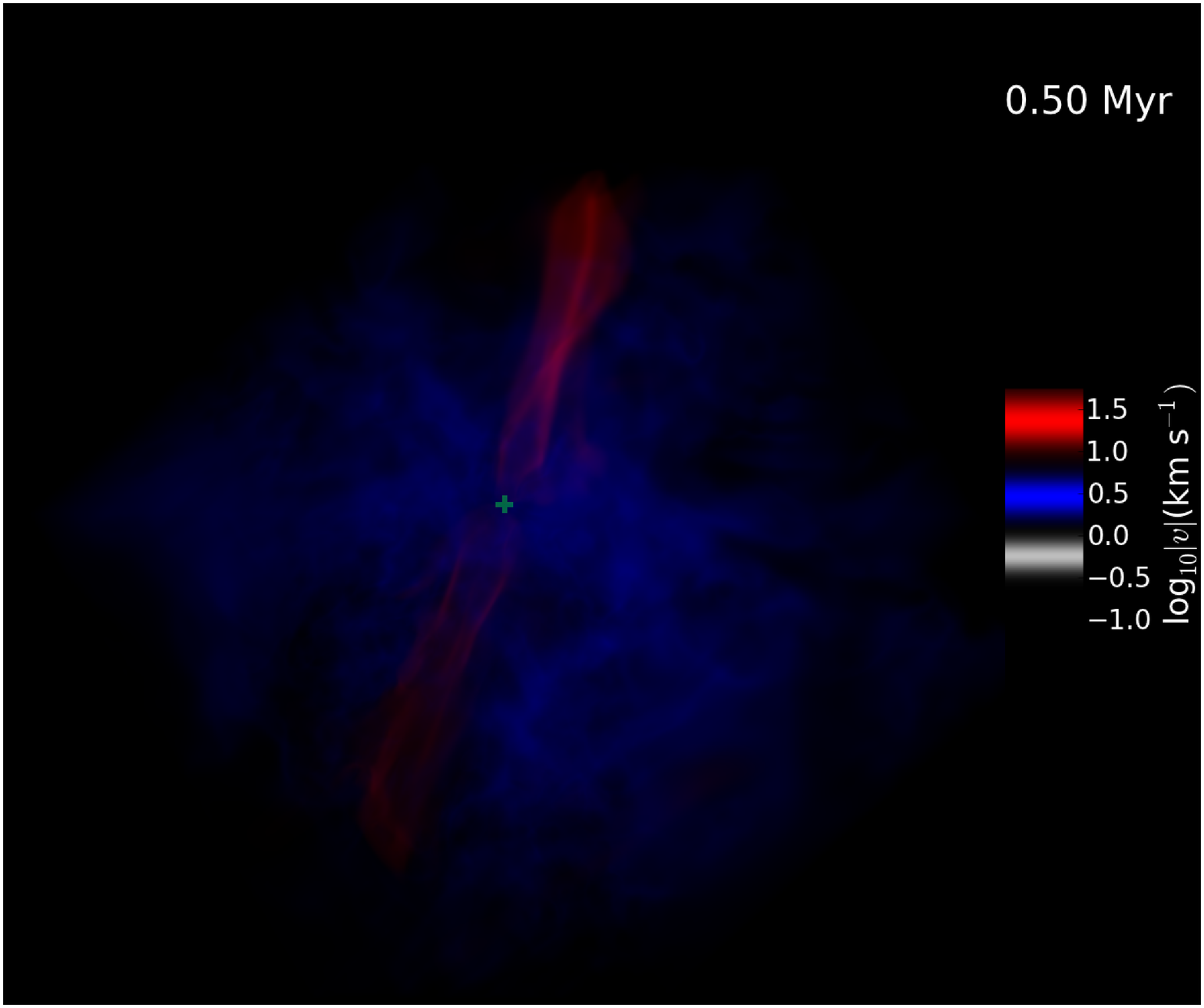}
\caption{Same as Figure \ref{VR_1} but for simulation th0.01. A movie is available showing the full time sequence.
\label{VR_01} }
\end{figure*}

The simulated outflows exhibit a variety of morphologies over time. Figures \ref{VR_1} and \ref{VR_01} show volume renderings of the gas velocity at six times for models with $\theta_0=0.1$ and  $\theta_0=0.01$. The initial turbulence in the core produces significant asymmetry between the upper and lower outflow lobes as well as asymmetry about the launching axis. Since the outflow itself is driven symmetrically, most of these differences arise from the interaction between the outflow and envelope gas. For example, the top, right panel of Figure  \ref{VR_01} shows that the lower lobe breaks out of the core earlier, while the upper lobe remains confined. Similar outflow asymmetries were also found in \citet{Offner11}.

At intermediate times the $\sim 30$ km s$^{-1}$ gas appears relatively similar in the two cases, despite the different launching angles. This is because the outflow successfully sweeps up and entrains core material in both cases. In fact, much of the high-velocity (red) gas shown is entrained material, which has mixed with the launched material. The figures show that gas at wider angles moves systematically slower. Sometimes slower moving, less collimated material is attributed to a ``disk wind'' mechanism, wherein the outflowing material is launched at larger radii with slower velocities \citep{pelletier92}.  However, here there is only a single launching mechanism that gives rise to a continuum of outflow morphologies and properties.

At late times, when most of the core gas has been accreted or dispersed, little entrainment occurs and the opening angle differences are more apparent. For example, the last panel of Figure \ref{VR_01} shows a very collimated outflow compared to that in Figure \ref{VR_1}.

The evolution of the gas velocity in the surrounding material is quite striking. Most of the initial core turbulence decays within 0.2 Myr and the core gas has velocities $\lesssim 0.4$ km s$^{-1}$, which is uncolored in Figures  \ref{VR_1} and  \ref{VR_01}. However, the fraction of diffuse gas with velocities $2-3$ km s$^{-1}$ increases with time. This is generally warm ($T>50 K$) gas that has been shocked but not unbound by the outflows. This demonstrates the potential for outflows to drive turbulence in their surroundings, and we explore this more quantitatively in Section \ref{turb}.

\begin{figure}
\epsscale{1.2}
\plotone{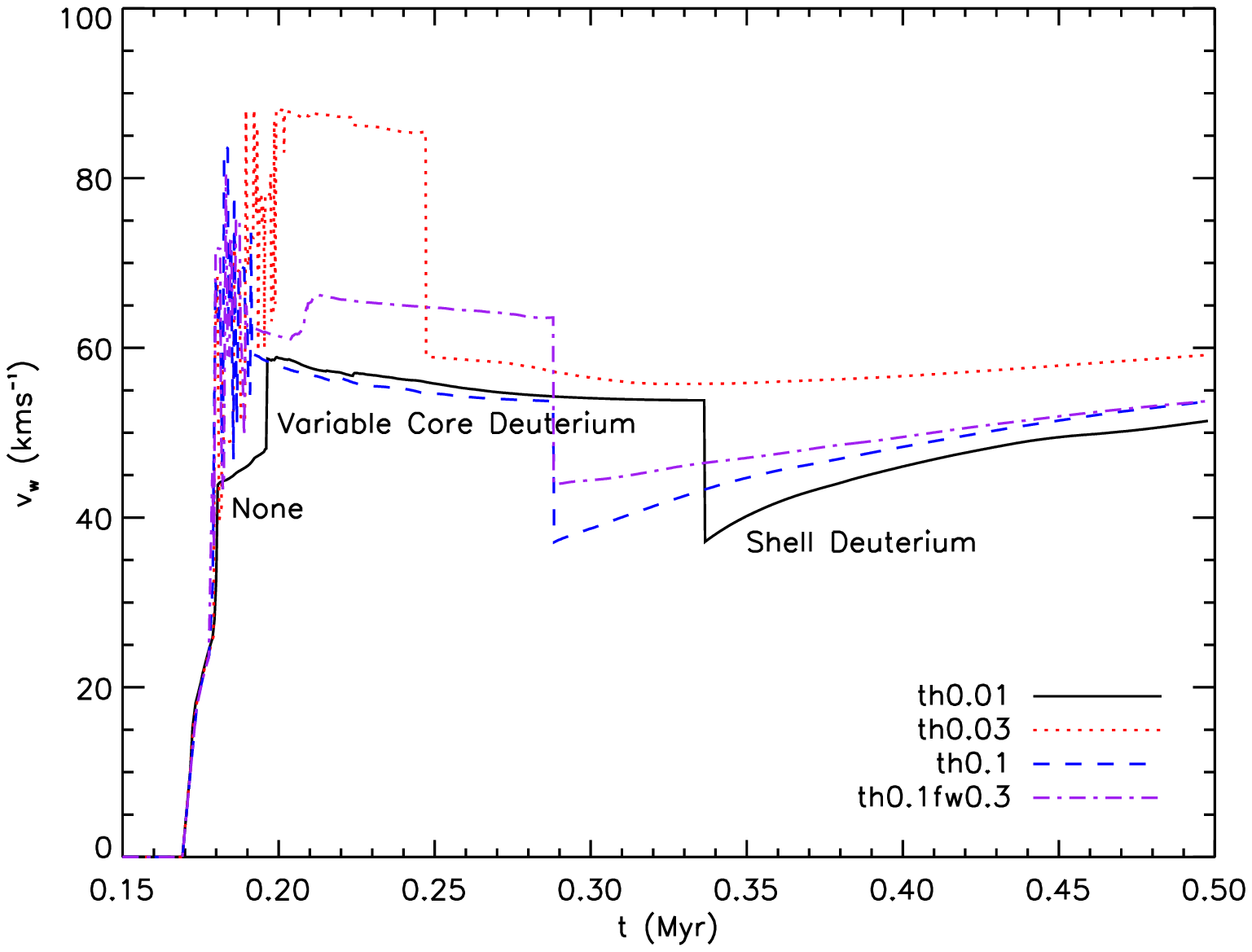}
\caption{Launching velocity, $v_{w} = f_v\sqrt{ GM_*/r_*}$, as a function of time for the four models. The protostellar evolutionary states are divided in to five different deuterium burning states, which depend on the temperature and density of the protostar: ``None" (no deuterium burning), ``Variable Core Deuterium", ``Steady Core Deuterium", ``Shell Deuterium", and ``ZAMS" (hydrogen burning commences). The duration of the states depends on the mass and accretion history of the protostar. For model th0.01, the ``Steady Core Deuterium" stage is brief and is not labeled.
\label{vwind} }
\end{figure}

\subsection{Mass Evolution}

The momentum distribution of the outflow and its interaction with the surrounding core gas have direct implications for the protostellar accretion rate and mass.  As discussed in \S\ref{methods}, we define the outflow launching velocity, $v_w$, as a function of the protostellar properties. Figure \ref{vwind} shows the launching velocity versus time for the four models. The curve discontinuities correspond to different interior stellar states with different amounts of deuterium burning; three of these states are indicated for model th0.01 (see the Appendix in \citealt{Offner09} for more details). At early times the stellar state is somewhat sensitive to accretion rate variations. Jumps in the wind launching rate are due to changes in the stellar radius, which depends upon the deuterium burning state. However, after the first $\sim$0.05 Myr, the accretion details have little impact. The gradual increase in $v_w$ from $t\sim0.3-0.5\,$Myr is due to the slow rise in $M_*$. At late times, the model $v_w$ converge to similar values because the protostellar masses and radii are comparable. More comprehensive stellar evolution models including variable accretion likewise predict that early accretion creates minimal scatter in protostellar properties after $\sim1\,$Myr (e.g., \citealt{hosokawa11}).

We find that the outflow evolution and gas entrainment is not linear in $\theta_0$ even though the simulations begin with identical initial core masses and turbulent velocity distributions. Some of the differences that result are due to differences in the early protostellar accretion, which depend on the extent the wind launching impacts the accretion flow. 
%For example, the run with the narrowly collimated angle undergoes only one instance of secondary fragmentation.
In the wider angle case, the infalling material interacts more with the outflow, leading to additional fragmentation. The widest angle wind run undergoes the most fragmentation episodes ($>4$), while the narrowest angle run experiences only one.  If different numbers of small fragments accrete onto the protostar, this will translate into small changes in the primary mass and later evolution.

The early phase of accretion and disk building lasts for $\sim$ 0.06 Myr. During this time streams of gas feeding the inner region deposit material with different angular momenta, producing significant changes in disk structure and total angular momentum. After this time, the gas settles into a stable Keplerian accretion disk of radius $\sim$200 AU. We expect this size to be an upper limit since magnetic fields are not included \citep{commercon11,seifried11,myers13,krumholz13}. 

Figure \ref{column} shows the column density distribution at different times for run th0.01. The other runs show a similar evolutionary progression. As shown in Figure \ref{column}, the outflow breaks out of the core around 0.2 Myr. The outflow broadens and entrains additional gas from $\sim$ 0.2-0.3 Myr.
By 0.35 Myr, the initial core has been almost completely disrupted or accreted. Thereafter, the protostar accretes the remaining turbulent, cold gas (green), more slowly. 

\begin{figure}
\epsscale{1.2}
\plotone{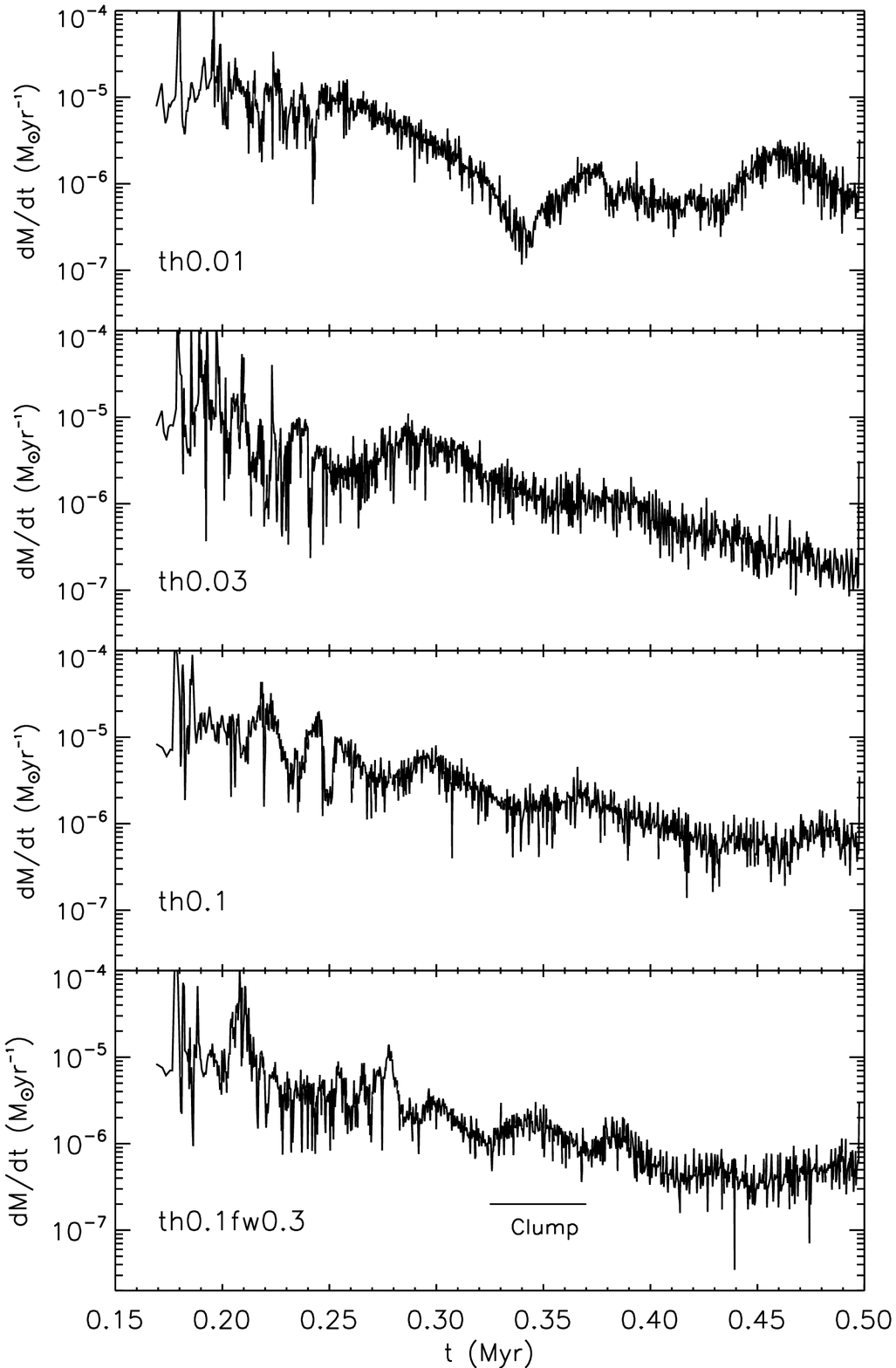}
\caption{Accretion rate as a function of time for the four runs.
\label{mdot} }
\end{figure}

\begin{figure*}
\epsscale{0.85}
\plotone{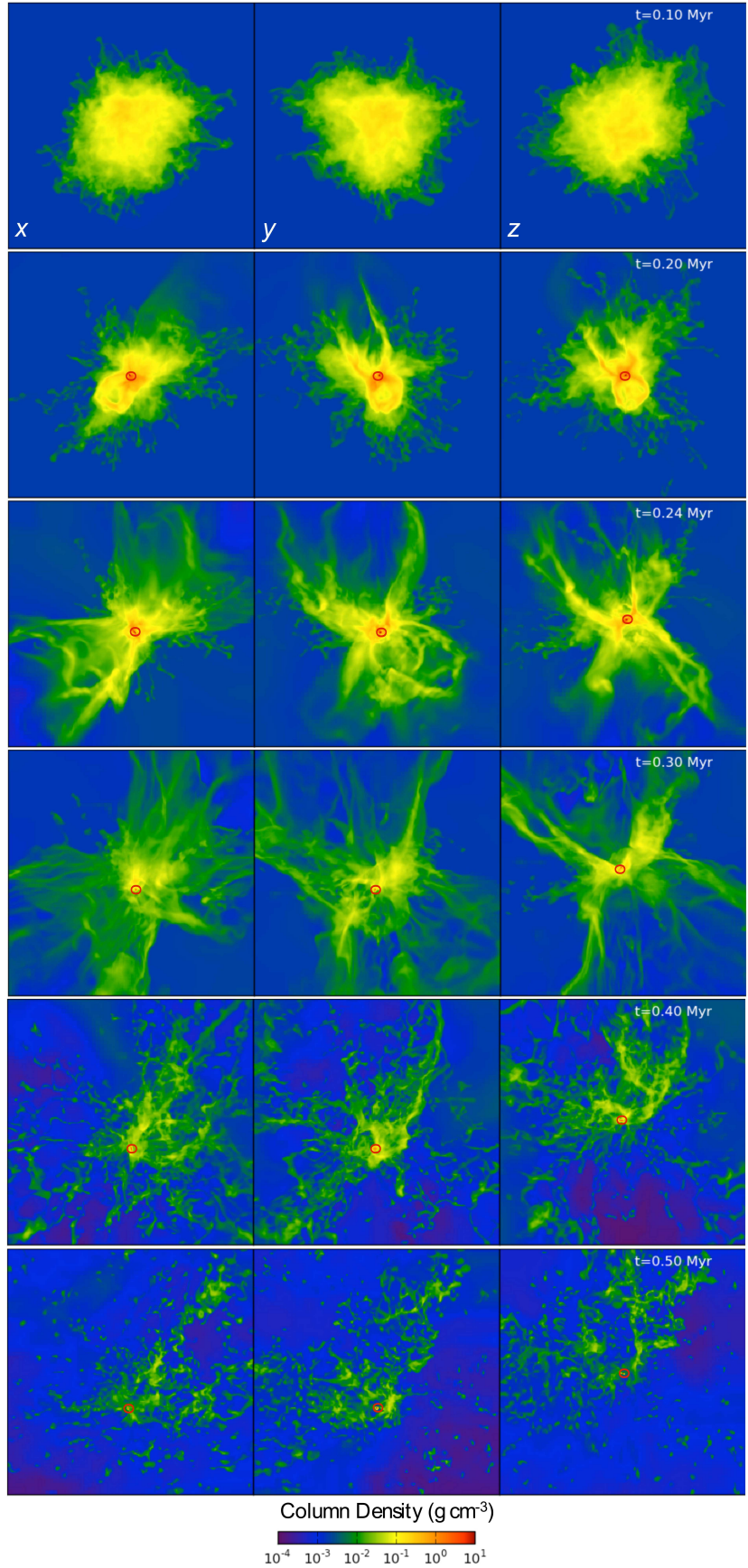}
\caption{Column density for six output times for run th0.01 for projections along the three cardinal axes. The output time is indicated in the upper right. The protostar location is denoted by the red circle.
\label{column} }
\end{figure*}

Figure \ref{mdot} shows the accretion rates for the four runs as a function of time.  The outputs have slightly different but comparable output intervals, with a median spacing of $\sim 300$ yr. All cases exhibit a decline in the mean accretion rate by more than an order of magnitude over the 0.3 Myr accretion time. Variability is largest at early times, which corresponds to the period of disk building and clumpy accretion. A decline corresponding to the disruption of the envelope is clearly visible in the accretion history of th0.01 (top panel in Figure \ref{mdot}) and in the others to a lesser degree. Some features in the average accretion, which span 0.03-0.05 Myr, correspond to periods when the protostar accretes a clump of material that is falling inward. An example of this is marked on the bottom panel of Figure \ref{mdot}. The interaction between the outflow and the envelope creates a significant amount of clumpiness in the residual cold gas as shown in Figure \ref{column}.

At late times, the accretion disk is stable and accretion variability generally remains within a factor of $\lesssim 2$ of the average accretion rate. Radiative heating acts to suppress large scale instability, and the accretion rate from the envelope onto the disk is sufficiently low that we do not expect global instability to occur \citep{kratter10,offner10}. However, the simulations do not have sufficient resolution to resolve small planet-size clumps if they occur (e.g., \citealt{vorobyov13,tsukamoto13}).

\begin{figure}
\epsscale{1.2}
%SSRO Should the top plot be linear
\plotone{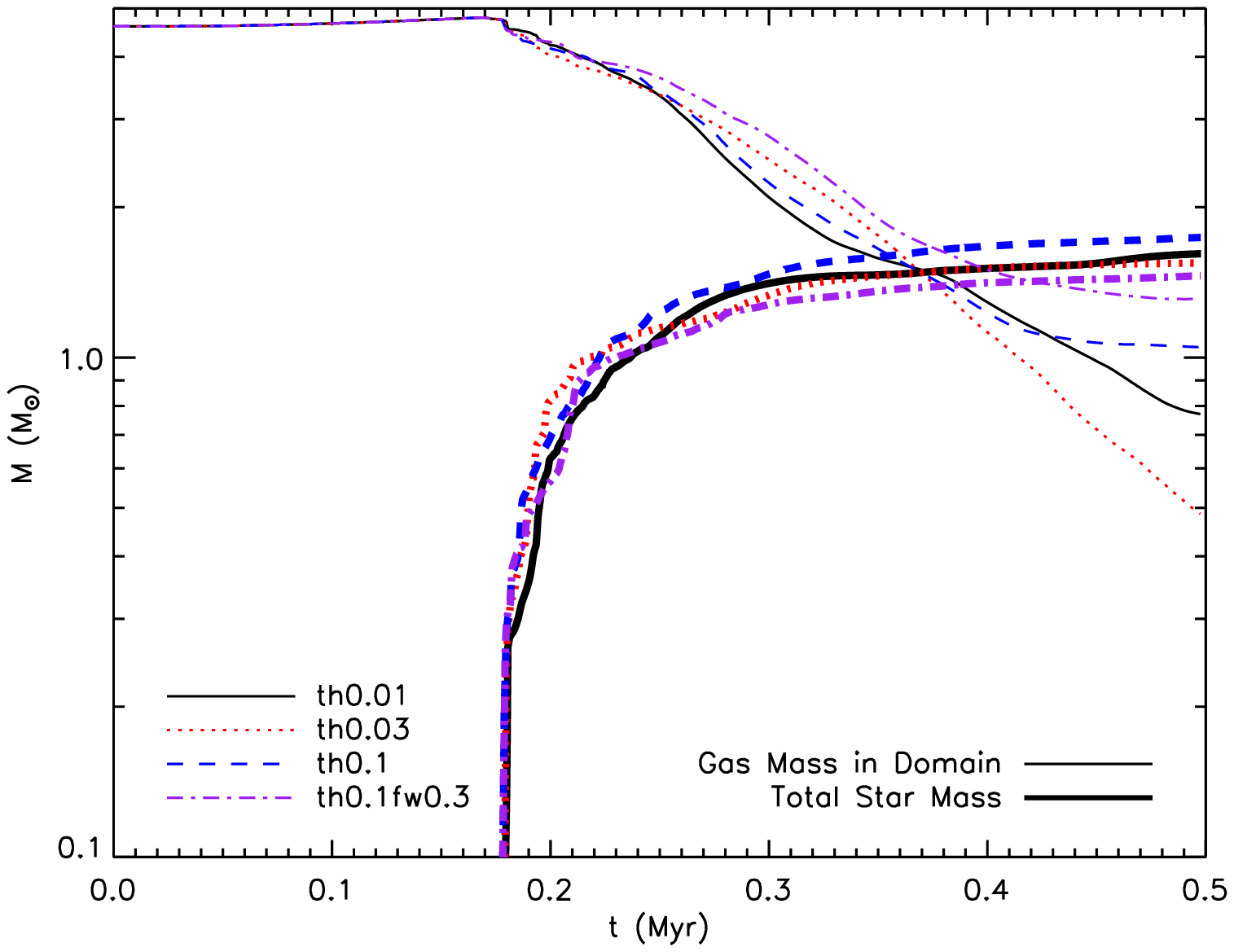}
\vspace{-0.1in}
\plotone{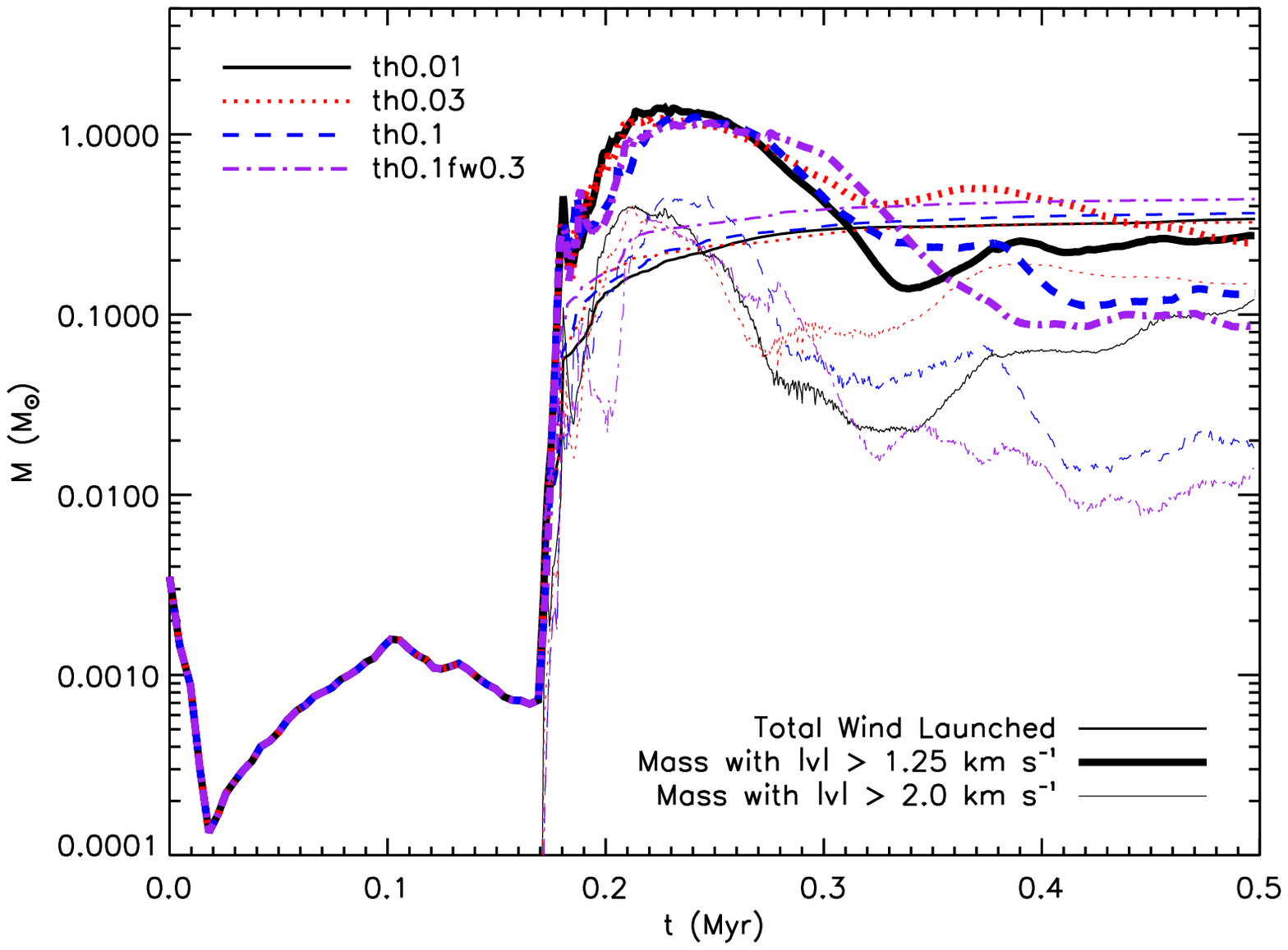}
\caption{Top: total gas in the domain and the stellar mass as a function of time (thick) for the four runs. Bottom: launched outflow mass, the mass with $lv| >$ 1.25 km s$^{-1}$ (thick), and the total mass with $|v| > 2$ km s$^{-1}$ (thin). 
\label{massvtime} }
\end{figure}

%outflow masses, accretion rates, and launching velocities as a function of time for all three models.
The differences between the runs is illustrated by Figures \ref{massvtime} and \ref{stacked}, which show the evolution of the protostellar masses, gas mass, and outflow mass as a function of time. Neither the evolution of protostellar mass nor the amount of high-velocity gas is monotonic with opening angle. This is partially due to differences in the early fragmentation and evolution and partially because of the initial turbulence within the core impacts the entrainment and evolution of the outflows. Small changes in the orientation of the outflow and the protostellar mass result in different results.

As expected, the run with the highest outflow efficiency has the lowest protostellar mass and the most mass on the domain at the end of the run. 
%This case also has less gas with velocities above $\sim1$ km s$^{-1}$, 
The run that has $\theta_0=0.1$, th0.1, also exhibits less mass at high-velocities because the gas is less collimated and more of the outflow momentum is distributed at wider angles with lower velocities. Run th0.1 also concludes with more gas mass remaining on the domain. This suggests that the wider angles winds in the prescription are not as efficient at unbinding and expelling core gas. Counterintuitively, the narrower opening angle cases, th0.01 and th0.03,  conclude with less gas on the domain. They have higher amounts of high-velocity gas even though the total mass launched is slightly lower. This demonstrates that narrowly collimated outflows can entrain and expel dense material. This is a different result than found by \citet{banerjee07}, who investigate a highly-collimated jet interacting with a magnetized medium. However, these authors adopted smooth, non-turbulent initial conditions; density inhomogeneities is an important component in coupling the outflow momentum and surrounding gas (e.g., \citealt{cunningham09,wang10}).

%plot_all_outflow_proop

\subsection{Protostellar Evolutionary Stage}

One of the main goals of observing individual protostars is to determine the evolutionary stage of the source. The earliest evolutionary period, Stage 0, during which most of the accretion occurs, is defined to be the length of time during which the envelope mass, $M_{\rm env}$, is greater than the protostellar mass, $m_p$ \citep{andre00,enoch08}. While $M_{\rm env}$ may be relatively easy to obtain from continuum emission measurements, the protostellar mass is nearly impossible to determine because protostellar masses cannot be directly measured.  In rare cases, high-resolution observations of  accretion disk properties can provide indirect estimates of the protostellar mass \citep{tobin12}, but in most cases the stage must be inferred through indirect means such as spectral energy distribution (SED) modeling \citep{whitney03a,robitaille07}. A more observationally convenient method of source classification involves using the effective bolometric temperature of the SED to separate colder, more extincted sources, which are allegedly younger, from warmer sources, which appear to have less surrounding gas and are thought to be older.  However, projection effects \citep{whitney03a,offner12} and non-monotonic accretion rates \citep{dunham12} can make younger sources appear older than they are and vice versa. For example, a source viewed along the outflow cavity will have less extinction along the line-of-sight and appear younger than it would if it were viewed through an edge-on accretion disk. Similarly, a source undergoing an accretion burst will appear brighter and may look more evolved. Comparing SEDs to analytic models for their evolutionary stage combined with some direct imaging can help improve age estimates, but  SED classes are often assigned for individual sources without imaging and are thus likely poorly correlated with evolutionary stage. 

In contrast, when protostars are treated as an ensemble, the individual errors associated with projection and variability may average out, allowing determinations of the stage lifetime \citep{evans09}.  Using the statistics of protostars in local regions, the Class 0 phase, which represents the earliest and highest accretion phase, is inferred to have a lifetime of $\sim$ 0.1 Myr \citep{enoch09,maury11}. \citet{evans09} estimated a combined Class 0 and Class I lifetime of  $\sim$ 0.5 Myr, which suggests a Class I lifetime of 0.3-0.4 Myr. However, these times are very uncertain because they depend upon an assumed disk lifetime, which is $2\pm1$ Myr.

Simulations provide one avenue for exploring the underlying Class and Stage lifetime. \citet{offner12} demonstrated that at early times, the stage inferred for simulated forming protostars on the basis of SED characterization and modeling is generally correct, although, inferred properties such as the envelope and protostellar mass could be incorrect by more than a factor of 2. \citet{machida13} inferred Stage 0 lifetimes ranging from $0.2-0.9 \times 10^5$ yr.\footnote{\citet{machida13} refer to this as the ``Class 0 lifetime", but because their definition is based on the protostellar mass and {\it not} on SED characteristics such as the bolometric temperature or spectral slope, the times they report are more accurately the {\it Stage} lifetimes. }

\begin{figure}
\epsscale{1.15}
\plotone{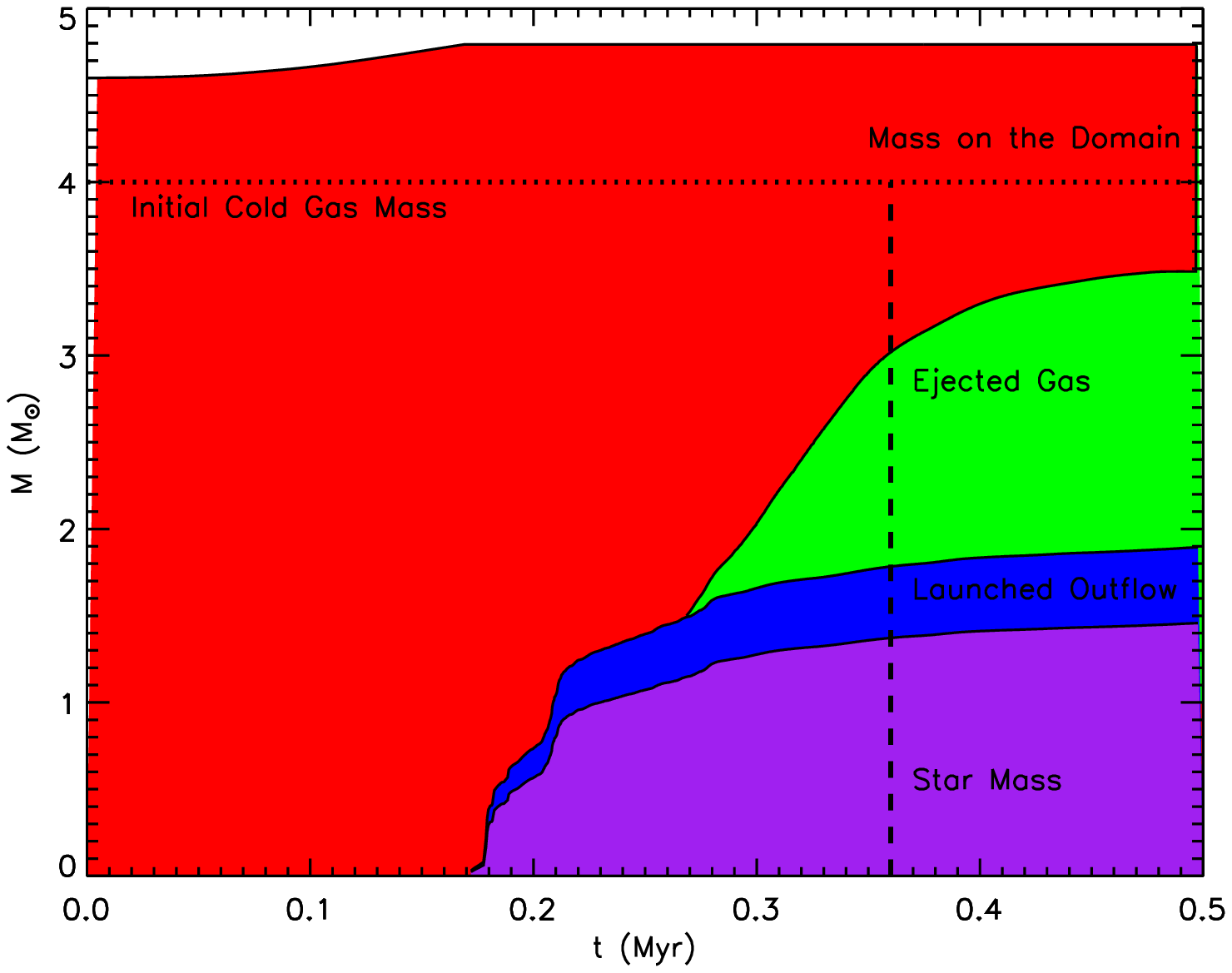}
\vspace{-0.1in}
\plotone{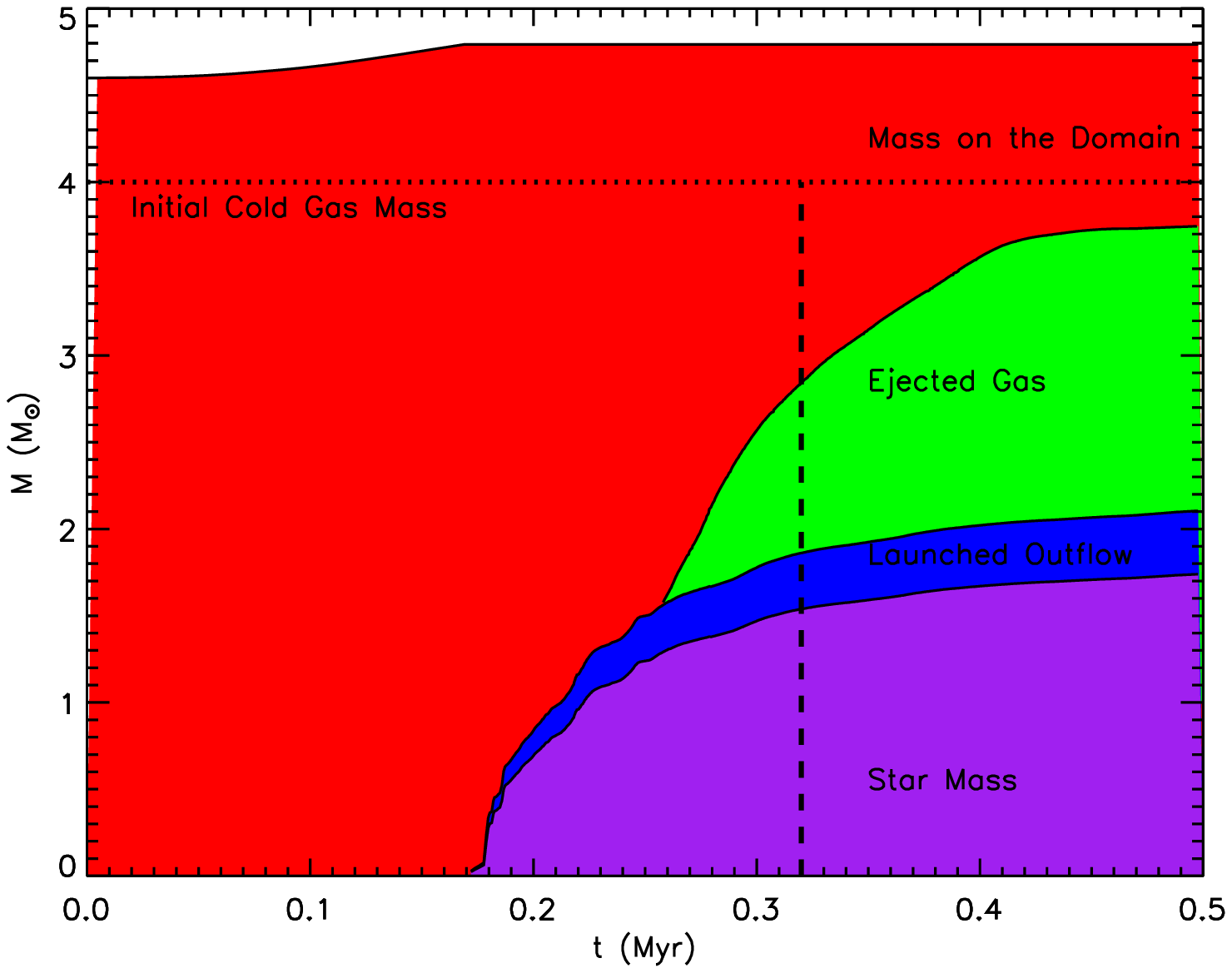}
\vspace{-0.1in}
\plotone{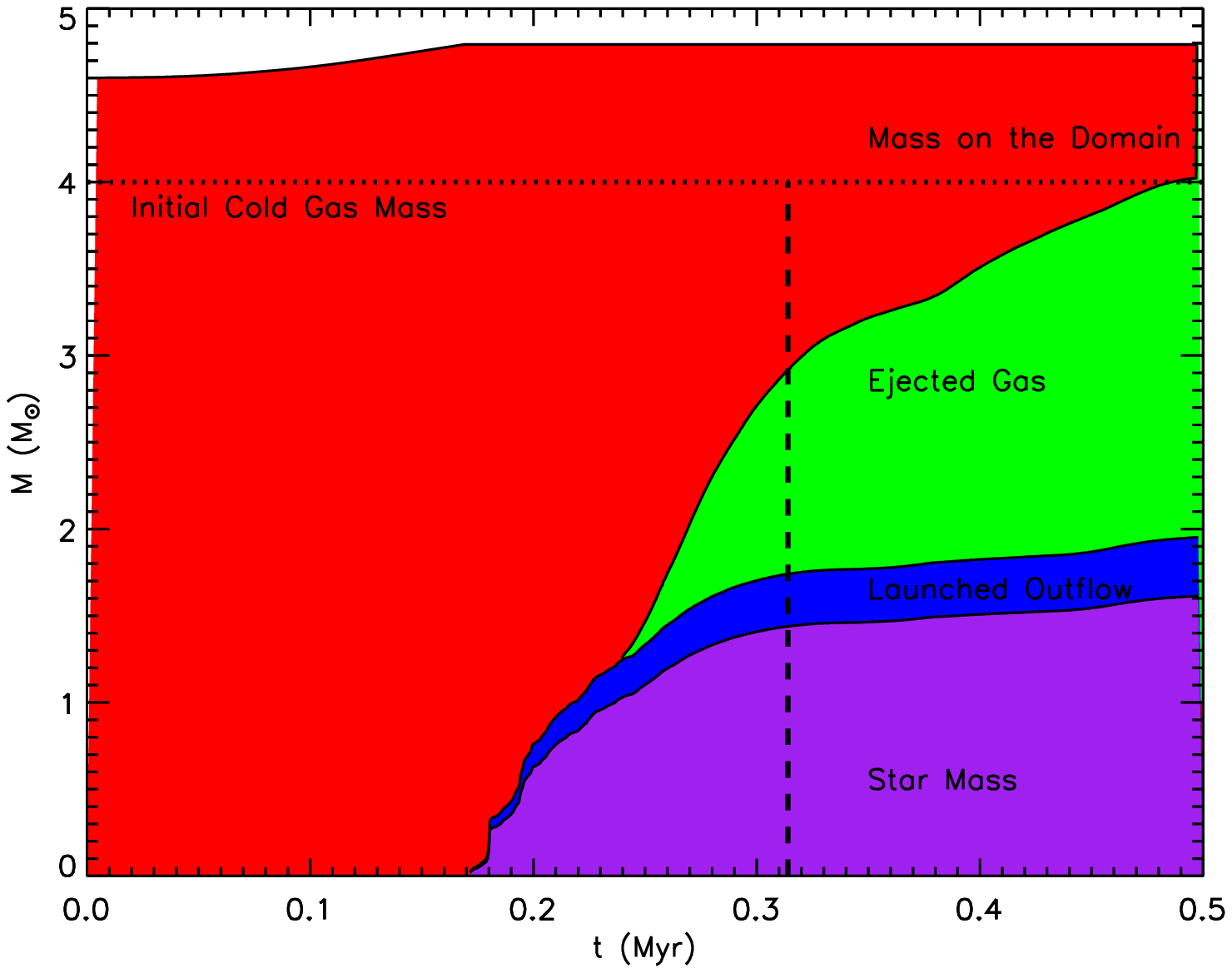}
\caption{Mass in the different components as a function of time for th0.1fw0.3 (top), th0.1 (middle) and th0.01 (bottom). The horizontal dotted line indicates the initial amount of 10 K gas (4.0 $\msun$), where the gas mass above the line is hot, low-density gas. The vertical dashed line indicates when the remaining envelope gas (defined as gas with $T < 50$K) is equal to the protostellar mass.
\label{stacked} }
\end{figure}

As plotted in Figure \ref{stacked} and indicated in Table \ref{simsum}, we find Stage 0 lifetimes of 0.14-0.23 Myr. These are slightly larger than found by \citet{machida13}, but they are comparable to estimates of the observed Class 0 lifetime. The range in values underscores that the length of the physical stage depends upon a range of initial properties, including outflow collimation, the initial turbulence and core properties. In calculations of isolated cores determining the core mass as a function of time is trivial. However, in simulations of clusters (e.g.,  \citealt{hansen12}), the envelope mass is connected with the cloud environment and may change with time due to additional accretion. Thus, in clustered conditions, simulations, like observations, must wrestle with the challenge of defining a ``core".

\subsection{Star Formation Efficiency}

Understanding the efficiency at which molecular gas turns into stars is an important theme in star formation, which has repercussions for the evolution of molecular clouds and the origin of the stellar IMF. On cloud scales, only a few percent of the gas turns into stars per dynamical time \citep{tan06,KandT07}. This appears to be a consequence of a combination of large scale supersonic turbulence, magnetic support, and stellar feedback. The efficiency of dense gas on the core scale is much higher. Since the decay time for turbulence is short and magnetic diffusion and reconnection reduce the field strength, the efficiency of dense gas is mainly limited by feedback and stellar multiplicity. Comparison between the stellar IMF and the observed distribution of core masses suggests an efficiency of $\epsilon \simeq \frac{1}{3}$ \citep{enoch08,alves07}, which naively implies that one-third of the gas in a core is converted into stellar mass. This is consistent with theoretical estimates \citep{matzner00}; although, if a core produces a binary or multiple star system the actual efficiency per core will be higher \citep{holman13}. 

Observationally, the derivation of the star formation efficiency is complicated by a number of factors including whether cores are bound, how many (if any) stars they form, and time-dependent effects. Consequently, $\epsilon \simeq \frac{1}{3}$ represents a fairly uncertain ensemble average of the efficiency.  Numerical simulations of isolated cores allow an unambiguous estimate of the star formation efficiency. In these calculations we find efficiencies of $\epsilon=$0.36-0.43 at 0.5 Myr as defined by the total mass in the protostar relative to the initial envelope mass (see Table \ref{simsum}). If we  ignore the remaining cold gas on the domain at 0.5 Myr, some of which may accrete onto the protostars if the calculations were run longer, we find $\epsilon=$0.41-0.51. These estimates suggest lower and upper limits on the efficiency at $t=\infty$, respectively.  These values are comparable to the estimated observed core efficiency, although this is very uncertain. Since each of these calculations formed only a single star, it makes sense that the efficiencies are larger than the average value obtained when comparing the core mass function to the IMF. 

%When star formation is considered for an entire molecular cloud, the rate at which molecular gas turns into stars is generally found to be very inefficient: SFR$_{\rm ff}\simeq 0.02$  \citep{KandT07}. 
The efficiency of star formation is more meaningful when considered together with some characteristic timescale, because, in principal, if there is no cloud dispersal mechanism then nearly all the gas will turn into stars on a sufficiently long timescale. The star formation rate (SFR) per free fall time is defined as the fraction of mass that turns into stars per free fall time \citep{krumholz05}. Here, the SFR$_{\rm ff}=m_{p,f}/ M_{\rm core}/(t_{\rm form}/t_{\rm ff})$, where $m_{p,f}$ is the protostellar mass at 0.5 Myr and $t_{\rm form}$ is the time over which the accretion occurred. Unimpeded global collapse gives SFR$_{\rm ff}$=1. As shown in Table \ref{simsum}, we find that the SFR$_{\rm ff}= 0.15- 0.18$ given that the free fall time of the initial core is $t_{\rm ff} = 0.134$ Myr.\footnote{If the initial starless phase in the calculation is included then the SFR$_{\rm ff} \simeq 0.1$.} %0.10-0.12$.
These calculations, which only consider an isolated core, yield a fairly low SFR. This implies that a combination of outflow feedback and core turbulence can contribute an order of magnitude to the global star formation inefficiency.
%2.5 \times 10^{-19}$ g cm$^{l-3}$, the star formation rate per free fall time is 0.13 

%plot_all_outflow_prop.eps
%SSRO should plot Cold gas mass

\subsection{Turbulence}\label{turb}

\subsubsection{Velocity Dispersion}

A number of authors have investigated the ability of outflows to contribute to the global energy budget in numerical simulation of clouds \citep{nakamura07,nakamura11,carroll09,wang10,hansen12}. However, little attention has been devoted to the study of outflow-driven turbulence on core scales, which we address here.

Figures \ref{VR_1} and \ref{VR_01} qualitatively demonstrate that the outflows in these simulation successfully drive turbulence on core scales. 
Figure \ref{sol} shows the root-mean-squared gas velocity dispersion, $v_{\rm rms}$, as a function of time for  the three runs. At early times, $v_{\rm rms}$ declines due to turbulent dissipation until the protostar forms.  Once the outflow is launched, $v_{\rm rms}$  increases non-monotonically as the high-velocity outflow gas interacts with the surrounding dense gas. The largest changes occur during the phase in which the outflow has not fully broken out of the core. At late times the velocities reach a quasi-steady state of about twice the initial velocity dispersion. We conclude that a single outflow is sufficient to offset the turbulent decay on sub-pc scales and can inject significant energy throughout the local region.

\subsection{Solenoidal and Compressive Motions}

Any velocity field can be deconstructed into two orthogonal components: a compressive field (${\bf \curl}  {\bf v}=0$) and a solenoidal field (${\bf \div} {\bf v} = 0$). Physically, these components indicate what fraction of the motions are ``squeezing" versus ``stirring." The compressive mode acts in a similar sense to gravitation in that it forces gas to higher densities. For star formation, the fraction of compressive motions may have bearing on the gas density and column density distributions \citep{federrath10}.

 In order to compute the solenoidal and compressive modes, we flatten the AMR data to a fixed $256^3$ grid (level 2). This allows us to easily perform Fourier Transforms of the data. Although this means that we neglect smaller scale motions, we find that the results are not significantly different if we adopt $128^3$ resolution instead, which implies that the larger scale modes dominate the totals.

\begin{figure*}
\epsscale{0.41}
\hspace{-0.25in}
\plotone{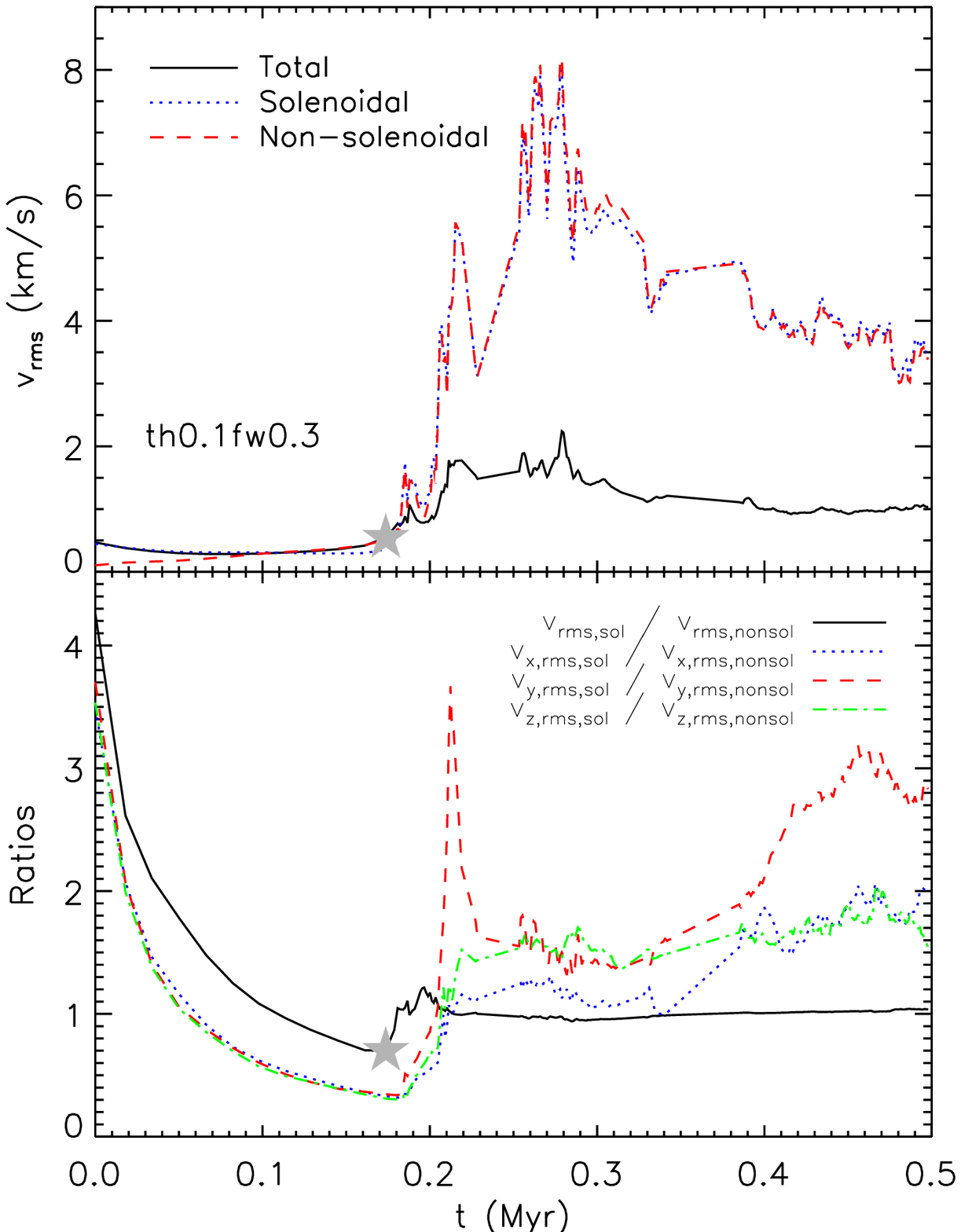}
\hspace{-0.25in}
\plotone{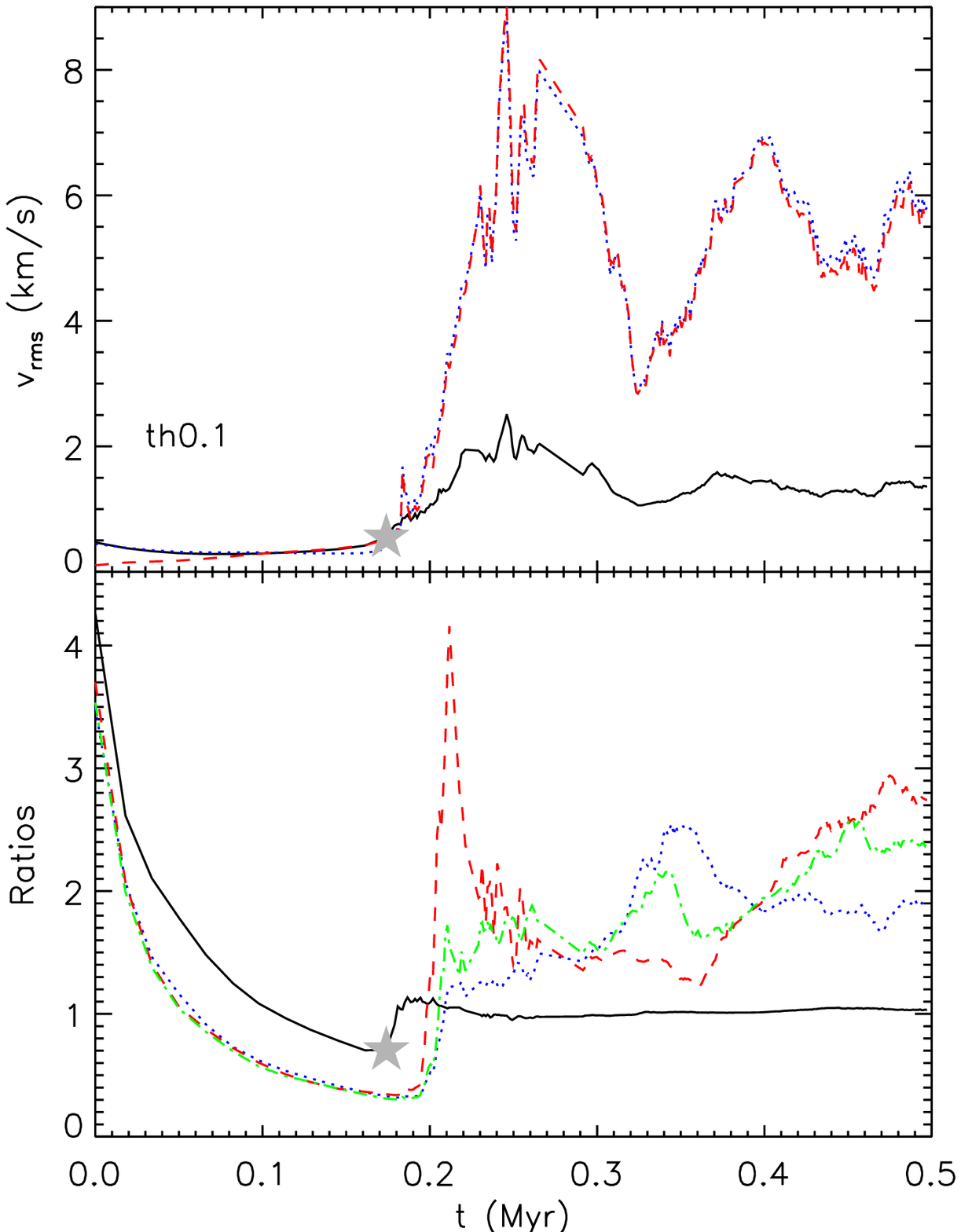}
\hspace{-0.25in}
\plotone{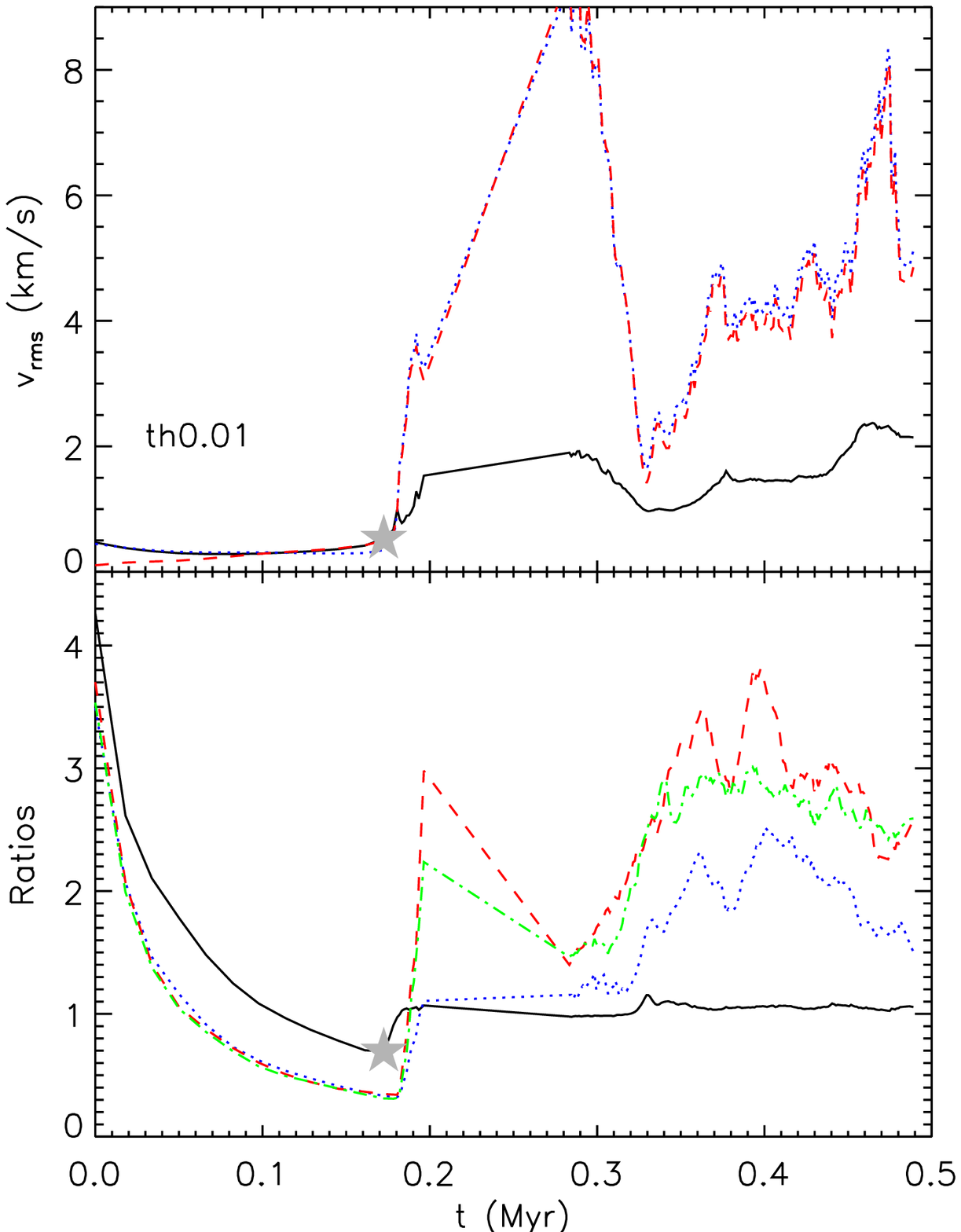}
\caption{ Top: mass-weighted total (solid), solenoidal(doted) and non-solenoidal(dashed) 3D velocity dispersion as a function of time. Bottom: Ratio of the solenoidal to non-solenoidal velocity fields for the mass-weighted 3D velocity dispersion (solid), the x-velocity component(dotted), the y-velocity component (dashed), and the z-velocity component(dot-dashed). The runs th0.1fw0.3, th0.1, th0.01 are shown from left to right. The gray star marks the formation time of the protostar. \label{sol} }
\end{figure*}

In these simulations the turbulence is initialized using a purely solenoidal random field. During the starless phase of the simulation, during which the turbulence decays and gravitational contraction begins, the solenoidal motions decline until compressive motions dominate  (bottom panels of Figure \ref{sol}). The launching of the outflow increases the solenoidal component, essentially by reducing gravitational collapse and creating circulation of material to larger scales. After 0.1 Myr, the ratio of the solenoidal to compressive velocity dispersion approaches one and the modes appear to be in equipartition.  Some of the compressive motion is due to ongoing gravitational contraction, as can be seen by the changing ratio for $t\lesssim 0.2$. This suggests that protostellar outflows inject turbulent motions that are more solenoidal than compressive.

The bottom panel of Figure \ref{sol} shows the ratio of the  solenoidal to compressive rms velocity along the three cardinal directions. The details of the distribution depend upon the orientation of the outflow with respect to the view.\footnote{Note that the individual ratios do not add to one because 
%\begin{equation}
 $\frac{ \sqrt{v_{x, {\rm rms},s}^2 + v_{y, {\rm rms},s}^2+ v_{z,{\rm rms},s}^2}}{\sqrt{v_{x,{\rm rms},c}^2 + v_{y,rms,c}^2+ v_{z,rms,c}^2}}  \neq \Sigma_i \frac{v_{i, {\rm rms}, s}}{v_{i,{\rm rms}, c}}.
$%\end{equation}
}
The top panel of Figure \ref{sol} shows the individual $v_{\rm rms}$ components of the solenoidal and compressive motions. These are higher than the total $v_{\rm rms}$ because ${\bf v}_{\rm tot} = {\bf v}_{s} + {\bf v}_c$, which means that at any given position $|{\bf v}_c|$ or $|{\bf v}_s|$ can be much greater than ${\bf v}$.
  
 \subsection{Core and Clump Turbulence Comparison}
  
\begin{figure}
\epsscale{1.15}
\plotone{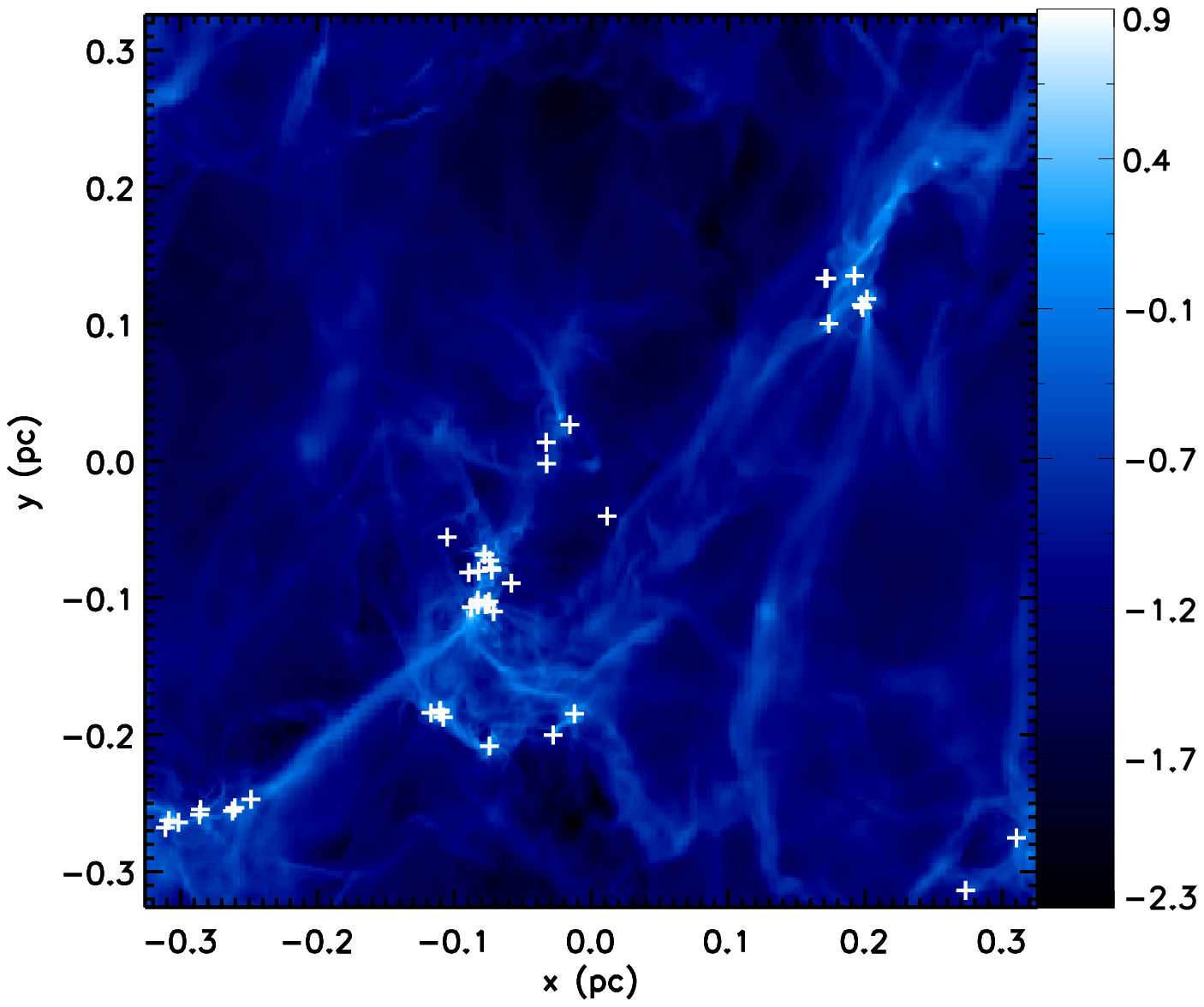}
\epsscale{1.1}
\plotone{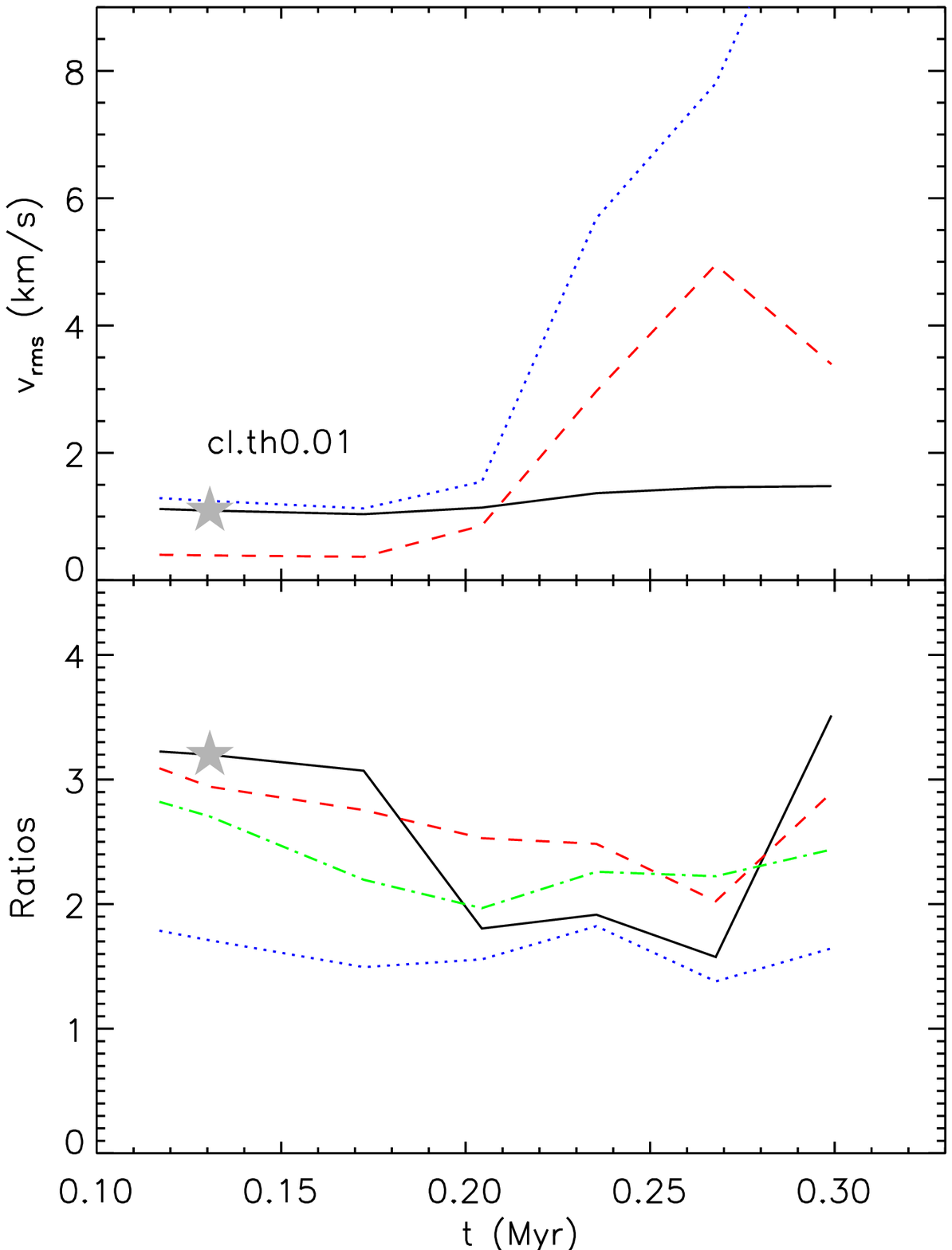}
\caption{ Top: Log of gas column density (g cm$^{-2}$) at time 0.3 Myr for run cl.th0.01. The white crosses mark the locations of stars. Bottom: Same as in Figure \ref{sol}, but for run cl.th0.01. The gray stars mark the formation time of the first protostar. \label{solcluster} }
\end{figure}

 In order to compare with turbulent outflow driving in a clustered environment, we analyze turbulent properties in a larger simulation, cl.th0.01, which is forming a cluster of stars (see Table \ref{simprop}). The clump simulation adopts periodic boundary conditions to model a $\sim1$ pc piece of a larger molecular cloud. Consequently,  a number of protostars form in fairly close proximity to one another, which is illustrated in the top panel of Figure \ref{solcluster}. Like the previously discussed core simulations, the initial turbulence is allowed to decay and the only kinematic input is from protostellar outflows.  Unlike the previous calculations the gas is not centrally condensed and gravitational motions are less ordered and more localized. Since gravity contributes to the compressive mode, we can use clth0.01 to examine the outflow driving in a  context where gravity is less dominant. We note that the cl.th0.01 simulation has periodic boundary conditions, and high-velocity material cannot leave the domain, which leads to an excess of turbulence in comparison to the case where outflowing material is allowed to escape the domain. As shown in the top panel of Figure \ref{solcluster}, the outflow driving is not directly discernible from the gas column density distribution.
 
The bottom panels of Figure \ref{solcluster} show $v_{\rm rms}$ and the solenoidal and compressive components as a function of time. As in the isolated case, the total dispersion and the dispersion of the solenoidal and compressive components increase once stars begin forming and launching outflows. However, the solenoidal and compressive modes are not as similar, and the outflows appear to drive much more solenoidal motion than compression. This could be because the cloud is not centrally condensed or globally collapsing, which is the case in the isolated core runs, and consequently, gravity contributes less compressive motion. This trend is illustrated by the solenoidal to compressive ratio, which is greater than unity at all times and in all three directions.
  
\section{CO Molecular Emission Maps}\label{obs}

In this section, we investigate the outflow morphology and velocity distribution inferred from emission in several observational tracers.
In order to generate synthetic emission maps, we post-process the simulations at selected intervals with {\sc radmc-3d} \citep{dullemond12}. We use the non-LTE Large Velocity Gradient approach \citep{shetty11}, which computes the level populations given input 3D density and temperature fields. We flatten the AMR data to a fixed 256$^3$ resolution ($dx=$0.001 pc). To interpolate over velocity jumps between grid cells we use ``doppler catching'' with $d_c=$0.025, which interpolates the velocity field such that velocity changes between points do not exceed 0.025 times the local line width. For the warm, high-velocity gas component, the thermal line width can exceed 1.4 km s$^{-1}$, so this small value is helpful for hot cells that border denser, cold gas. We adopt a constant ``microturbulence'' value of 0.05 km s$^{-1}$ in order to include sub-resolution line broadening caused by unresolved turbulence. We assume a constant molecular abundance except in the hot ($T> 900$K) where the molecules are assumed to have much lower abundances or be completely dissociated. The molecular excitation and collisional data are adopted from the Leiden atomic and molecular database \citep{schoier05}. We use the Rayleigh-Jeans approximation ($h\nu < k_BT$) of the Planck function to compute the brightness temperature corresponding to the output flux.\footnote{It is standard for observers to employ this approximation to obtain the observed brightness temperature even if the emission is not in the Rayleigh-Jeans limit.}

%\subsection{CO}

\begin{figure*}
\epsscale{1.2}
\plotone{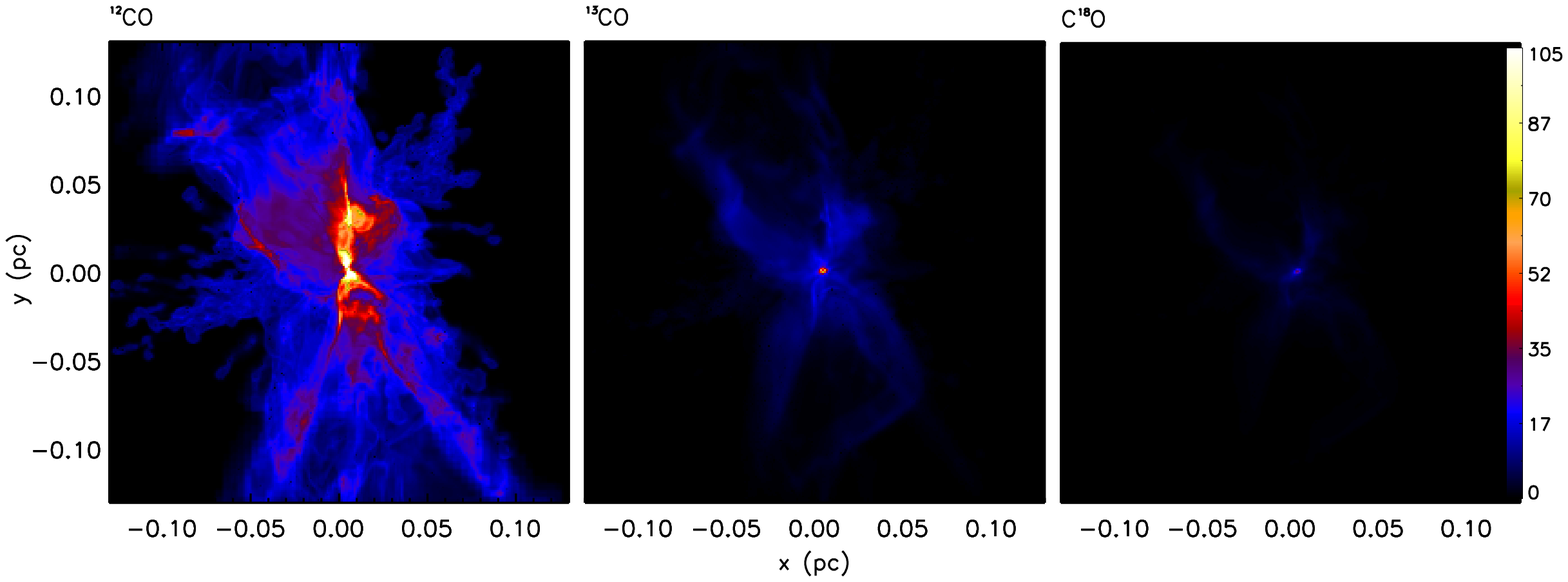}
\caption{Integrated emission (K km s$^{-1}$) for $^{12}$CO (1-0), $^{13}$CO (1-0), and C$^{18}$O (1-0) for a roughly edge-on view of th0.1fw0.3 at time 0.24 Myr when $M_* = 1 \msun$. The integrated emission is in K km s$^{-1}$. \label{COiso} } %This assumes Rayleight-JEans
\end{figure*}

As the second most abundant molecule in star-forming regions after H$_2$, CO is the most commonly used tracer for $n\sim10^2-10^3$ cm$^{-3}$ gas. CO is also useful as a tracer of the lower density, molecular component of protostellar outflows \citep{arce07}.  The CO isotopologues, $^{13}$CO and C$^{18}$O, have lower abundances and higher critical densities and so are useful tracers for examining slightly higher density gas, where $^{12}$CO is optically thick.
Here, we assume a constant CO abundance of ${\rm CO}/{\rm H_2} = 8.6 \times 10^{-5}$ for cells with temperatures below 900 K. Gas above this temperature is either strongly shocked, which may dissociate the CO or low-density, high-pressure confining gas, which was not part of the original molecular cloud core. We adopt $^{12}$CO/$^{13}$CO=62 and C$^{18}$O/H$_{2}$ =$1.7\times10^{-7}$. 

Figure \ref{COiso} shows the integrated emission for $^{12}$CO(1-0), $^{13}$CO(1-0), and C$^{18}$O(1-0). The $^{12}$CO emission is very bright and clearly traces the initial core gas as well as the outflow. Clumps and streams of gas can be seen being ejected from the inner region. $^{13}$CO only traces the outflow cavities and dense core center; very little of the initial core or ejected material is apparent. The C$^{18}$O picks out only the inner cavity structure, and the cavity appears more rounded than in the other two tracers.

\begin{figure*}
\epsscale{1.2}
\plotone{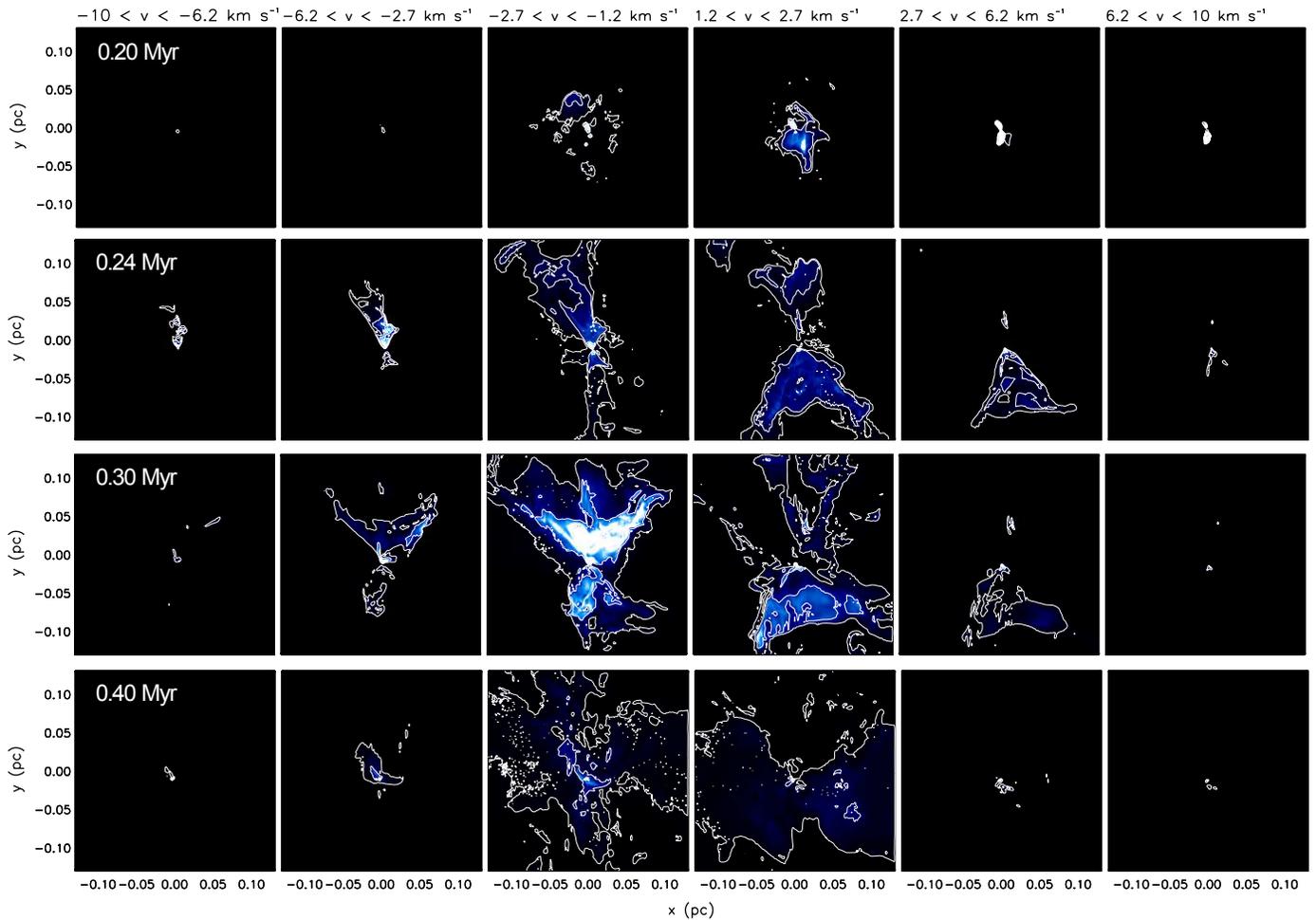}
%\plotone{plrc01250_0_90_allchan.eps}
%\plotone{plrc02640_0_90_allchan.eps}
%\plotone{plrc04970_0_90_allchan.eps}
%\plotone{plrc08390_0_90_allchan.eps}
\vspace{-0.25in}
\caption{Integrated $^{12}$CO(1-0) emission for different velocity channel ranges for the run th0.1fw0.3 at times 0.20, 0.24, 0.3, and 0.4 Myr. The simulation is viewed such that the outflow is titled $\sim 30$ degrees with respect to the line-of-sight.  The contours are shown for 0.5 K $\kms$, 5.0 K $\kms$, and 20.0 K $\kms$. \label{COchan} }
\end{figure*}

Figure \ref{COchan} shows the integrated $^{12}$CO(1-0) emission in a set of velocity channels for four different times. At the earliest time, the high-velocity gas is very localized. Once the outflow has broken free of the core, the emission takes on a distinctive ``v'' shape, which has been observed in a variety of observed outflows (e.g., IRAS3282,  \citealt{arce06}).  The outflow emission characteristics at $\gtrsim 0.3\,$Myr  are also similar to those reported by \citet{quillen05} who mapped NGC 1333 in $^{13}$CO and found cavity sizes of $\sim 0.1-0.2$ pc and velocity widths of a couple $\kms$.
%The different opening angles show similar morphologies at this time, but the amount of apparent collimation and shape of the cavity depends on the viewing angle with respect to the line of sight. 
%For more inclined views, the cavity can appear more rounded. 

The amount of gas that appears at higher velocities depends partly on the orientation of the outflow with respect to the line-of-sight. 
The figure shows a view in which the outflow is titled $\sim 30$ degrees with respect to the line-of-sight. At smaller angles of orientation, the outflow cavity appears more collimated. The emission with velocities close to 10 km s$^{-1}$ is mainly concentrated near the protostar.  Much higher velocity gas is present than appears in the maps because much of the highest velocity gas has temperatures exceeding the cutoff of $900$K. At the launch point, the wind is assumed to be ionized ($T=10^4$ K).  This gas quickly begins to cool and much of the high velocity gas is in fact around a few thousand Kelvin. In principle this hot, high-velocity gas may emit in CO if some of it is entrained and heated core gas, since CO does not dissociate until $\sim 5\times10^3$ K. However, given that the formation timescale of CO is much longer than the dynamical time of the outflow, we expect the CO abundance of this material to be much lower and the contribution to the total $^{12}$CO (1-0) emission from this hot molecular gas to be small. %Figure \ref{tilt} shows the outflow at $t=0.24$ Myr inclined by 45.

While there are a few instances of discrete clumps of material, the outflows do not exhibit the apparent episodicity exhibited by outflows like HH 46/47 \citep{arce13}. Although the simulated protostars do experience some accretion variability, it is generally less than a factor of a few over consecutive time steps. Observed episodic clumps may be created by a very high-velocity jet component, which is also not well resolved here or is not reproducible with our simple outflow model.

Toward late times, the higher velocity gas is more diffuse and patchy. The outflow is not directly apparent in the channel maps, instead the emission traces the residual turbulent core gas. Whether this complex velocity structure is resolvable in observations depends on the total cloud velocity dispersion, which may hide small features, and the beam resolution.

\begin{figure}
\epsscale{1.2}
\plotone{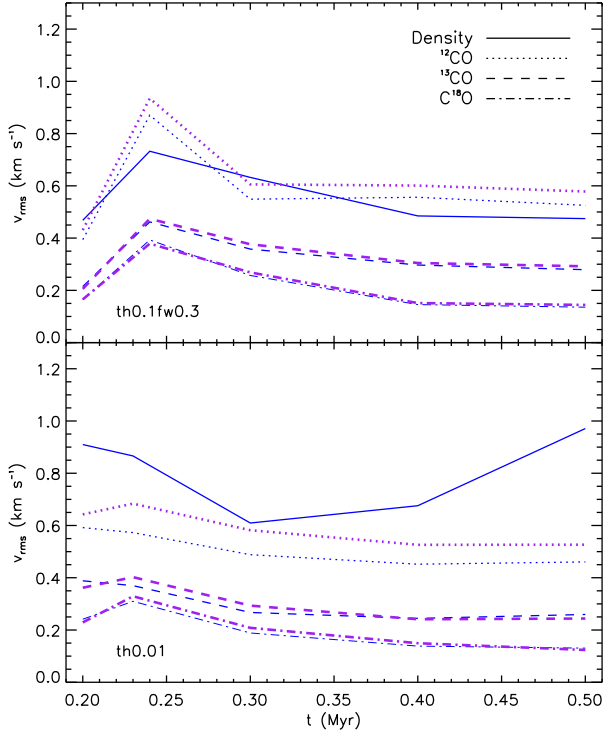}
\caption{ Average velocity second moment as a function of time for the simulations th0.1fw0.3 (top) and th0.01 (bottom) for a view along the $z$ axis. For the raw simulation data (solid), the velocities are density weighted; the moment is computed for gas velocities $-10 \leq v \leq 10$ km s$^{-1}$ and for gas temperatures  $T<500$ K.  For the  $^{12}$CO (dotted), $^{13}$CO (dashed), and C$^{18}$O (dot-dashed) emission %the velocities are weighted by the intensity and 
only channels with $T_B>0.1$ K are included in the moment calculation. The thick, purple lines show the average dispersion of the molecular emission convolved with a 5'' beam and where the source is assumed to be 250 pc away.
\label{vdispco} }
\end{figure}

The CO isotopologues are somewhat imperfect tracers of the velocity dispersion. Figure \ref{vdispco} shows the average second velocity moment as a function of time for two simulations computed directly from the simulated data and from the $^{12}$CO, $^{13}$CO, and C$^{18}$O emission.  For the raw simulation data, the second velocity moment is simply the mass-weighted root mean squared velocity. For the emission data, the average second moment is defined as the average of the map second moments:
\beq
\sigma=\frac{1}{N_{\sigma_{x,y} > 0}}\sum_{\sigma_{x,y} > 0} \left( \sqrt{ \frac{ \int T_B(x,y,v)[v-\bar V(x,y)]^2 dv}{\int T_B(x,y,v) dv}} \right), 
\eeq
where $T_B$ is the brightness temperature, $\bar V(x,y) = \int T_B(x,y,v) v dv/ \int T_B(x,y,v)dv$ is the first moment at location $(x,y)$ and $N_{\sigma_{x,y}> 0}$ is the number of pixels in the map with non-zero dispersion.  The same data is shown where the emission has been convolved with a 5'' Gaussian beam assuming that the source is 250 pc distant. This is analogous to an outflow in Perseus observed with the Combined Array for Research in Millimeter-wave Astronomy (CARMA). 

Generally, the velocity dispersion increases slightly as the outflow expands into the core and then decreases as the envelope is accreted or expelled (see also Figure \ref{sol}).  
%The velocity dispersions of the isotopologues are generally expected to follow $\sigma_{\rm C^{18}O}< \sigma_{\rm ^{13}CO}<\sigma_{\rm ^{12}CO}$ in observations of dense cores, since these tracers probe different densities and relative scales.  
%However, the tracer with the largest dispersion varies as a function of time. At early times, the infalling inner envelope and entrained, dense gas move most rapidly, so the C$^{18}$O has the highest average dispersion. By late times, the fraction of dense gas is small and much of the remaining gas consists of low-density material, which is best traced by $^{12}$CO. 
The velocity dispersions of the isotopologues are generally expected to follow $\sigma_{\rm C^{18}O}< \sigma_{\rm ^{13}CO}<\sigma_{\rm ^{12}CO}$, where the smaller line width tracers probe relatively smaller sizes and higher density gas. Here, we find that the  line widths vary as expected. Convolving the emission with a beam preserves the order of the line width but tends to make the dispersions slightly larger. None of the tracers track the mass-weighted velocity dispersion well.  The $^{12}$CO dispersion is the most similar, although the actual line-of-sight dispersion may be either larger or smaller than the observed value. The difference between the synthetic line width and the simulation velocity dispersion is mainly due to varying excitation, which is a function of the local temperature, density, and optical depth. Since we assume constant abundances for the CO isotopologues, we expect that in this comparison the intensity of the synthetic emission maps  is better correlated with the underlying gas density than would be true for actual observations. 

\begin{figure*}
\epsscale{1.1}
\plotone{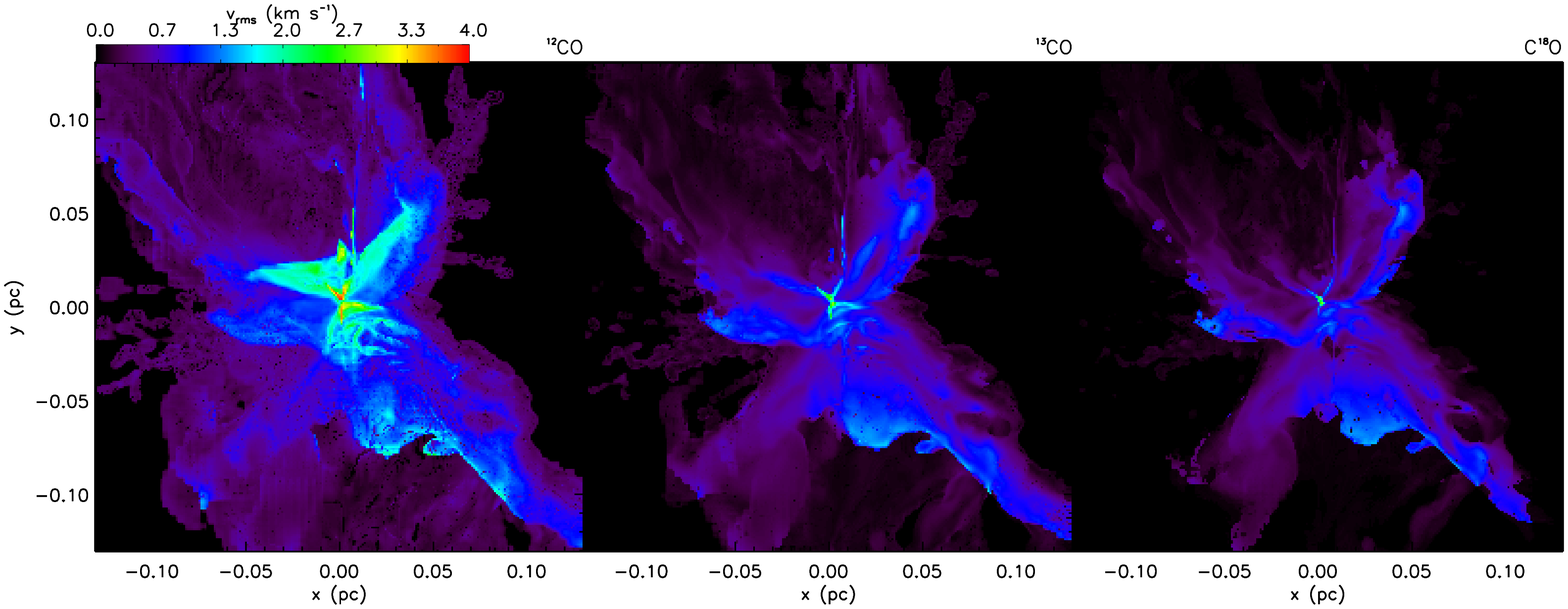}
\plotone{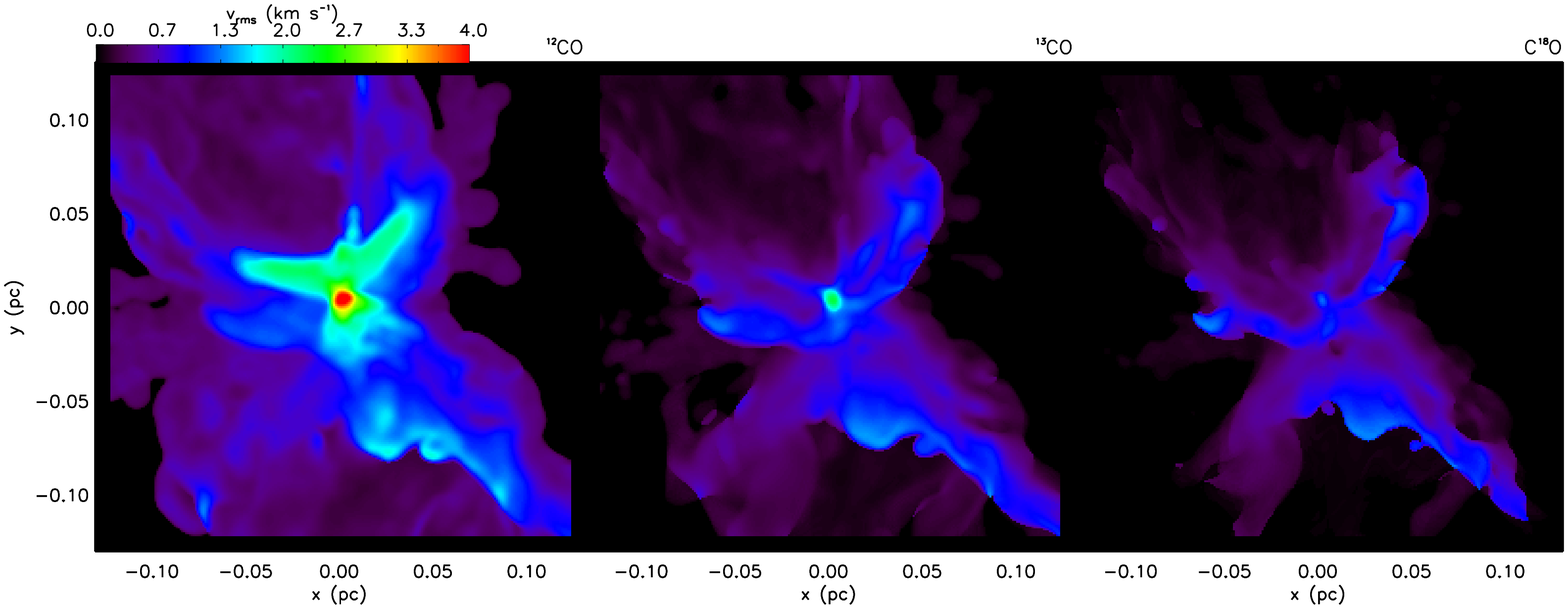}
\caption{Top: Map of the velocity second moment for the $^{12}$CO (left), $^{13}$CO (middle), and C$^{18}$O (right) emission  computed along the $z$ axis at 0.3 Myr for th0.01. Bottom: Map of the velocity second moment  where the emission has first been convolved with a 5'' beam where the source is assumed to be 250 pc away.
\label{vdispcomap} }
\end{figure*}

Figure \ref{vdispcomap} shows a map of the second moment for each of the isotopologues at one simulation snapshot. The dark areas in the plot indicate sightlines with no cold, dense gas. The $^{12}$CO traces the outflow best, and clumps of low-density, high-velocity gas are apparent in the upper outflow cavity, which is distinctly V-shaped, and on the left side. The other two tracers highlight more of the wider angle material, which is part of the outer outflow cavity. This material has second-moments of $\sim$1-2 km s$^{-1}$. The C$^{18}$O probes a smaller fraction of the gas, but its emission and estimated velocity dispersion still strongly resemble that of $^{13}$CO.

%but it does highlight some of the rotating material near the protostar. This material is on a larger scale than the actual accretion disk, which spans only a couple pixels in the image map. In the convolved image, the rotating gas is smeared out over the beam area.
%As the critical density and optical depth of the tracer increases, the filling fraction in the plot decreases. The highest dispersions are located near the protostar, along the cavity walls, and at the edges of the expelled gas.

%SSRO different inclination. to show higher velocity gas?

%\subsection{HCO$^{+}$}

%Should an image in this tracer be included here?

\section{Summary}\label{summary}

We perform a set of radiation-hydrodynamic simulations of isolated forming protostars with a model for outflow launching in order to investigate outflow evolution, turbulent driving, and star formation efficiency.  Our outflow model is based on the X-wind magnetized wind model, in which the outflow rate is coupled to the protostellar accretion rate and evolution. The gas temperature is calculated using flux-limited diffusion including radiative feedback from the forming protostar and atomic cooling at high temperatures. In the simulations, we vary the outflow collimation angle from $\theta=0.01-0.1$ and find that even well-collimated outflows  effectively entrain gas and drive turbulence.  Due to the inclusion of radiative heating, after the early infall phase, accretion onto the protostar is generally smooth and only varies by a factor of a few on timescales of $\sim 300$ yr. Consequently, while the gas accretion and outflow launching is somewhat variable, the outflow does not undergo large episodic bursts.

 The final turbulent velocity dispersion is about twice the initial value, which demonstrates that an individual outflow easily replenishes turbulent motions on $\sim$0.1 pc scales.  Although the initial turbulence in the simulations is purely solenoidal, we find that the resulting gas motions are approximately equally solenoidal and compressive. Some of the gas compression is due to gravitational collapse, so it appears that outflows drive motions that are predominantly solenoidal rather than compressive. We confirm this conclusion by analyzing an additional simulation of a cluster of forming protostars, which is not centrally condensed.  

We post-process the simulations with {\sc radmc-3d} to produce synthetic molecular line emission maps in $^{12}$CO, $^{13}$CO, and C$^{18}$O. The  emission morphologies of the simulated outflow cavities appear similar to observed outflows. The line width is anti-correlated with the tracer critical density, and as expected, $^{12}$CO is a better probe of the outflow gas. The effective dispersions of the different tracers are similar but not well-correlated with the actual dispersion computed using the 3D density information.  We find that the results are qualitatively similar if the emission is convolved with 5'' beam, although the inferred dispersions are larger.
%At early times, while the outflow punches through the core, the higher critical density tracers probe the outflow best, while at late times, when the dense gas is depleted, the $^{12}$CO is a better tracer.   

In future work we plan to address outflow entrainment and turbulent driving including the effects of magnetic fields. Some numerical simulations can produce magnetically launched outflows self-consistently. However, either resolution is insufficient to probe a highly collimated jet component or the calculation is evolved for a very short time due to computational time constraints. Most magnetic calculations also assume perfect coupling between the magnetic field and gas (ideal MHD), and no 3D calculation includes all the necessary physics modeling magnetic diffusion and reconnection. More accurate and complete future studies are needed to fully understand outflow launching, entrainment, and turbulence.
%The role of these processes in outflow launching and entrainment is relatively unexplored.

\acknowledgements We thank Christoph Federrath for helpful discussions and an anonymous referee whose constructive comments improved the paper. The images and animations for Figures 3, 4 and 7 were made possible by {\it yt} \citep{turk11}.
Support for S.S.R.O. was provided by NASA through Hubble Fellowship grant HF-51311.01 
awarded by the Space Telescope Science Institute, which is operated by the Association of
Universities for Research in Astronomy, Inc., for NASA, under contract NAS 5-26555.  H.G.A. acknowledges support from his NSF CAREER award AST-0845619. The simulations were performed on the Yale University Omega machine; this work was supported in part by the facilities and staff of the Yale University Faculty of Arts and Sciences High Performance Computing Center.

\bibliography{outflowbib}
\bibliographystyle{apj}

\end{document}